\documentclass[longauth]{aa}  

\usepackage{graphicx}
\usepackage{txfonts}
\usepackage{hyperref}
\hypersetup{colorlinks,allcolors=magenta}

\usepackage{orcidlink}
\newcommand{\orcid}[1]{\orcidlink{#1}}

\newcommand{\emm}[1]{\ensuremath{#1}}   
\newcommand{\emr}[1]{\emm{\mathrm{#1}}} 
\newcommand{\unit}[1]{\emm{\, \emr{#1}}}

\newcommand{\Jone}{\mbox{(1--0)}}
\newcommand{\Jtwo}{\mbox{(2--1)}}

\newcommand{\kms}   {\unit{km\,s^{-1}}}

\newcommand{\pc}    {\unit{\pc}}

\def \arcsec {{\rm $^{\prime\prime}$}}

\def \fdense {{$f_\mathrm{dense}$}}
\def \sfedense {{SFE$_\mathrm{dense}$}}
\def \sigmamol {{$\Sigma_\mathrm{mol}$}}

\begin{document} 

   \title{The properties and kinematics of HCN emission across the closest starburst galaxy NGC~253 observed with ALMA}
   
    \titlerunning{Dense gas in NGC~253}
   \authorrunning{Be\v{s}li\'c}


   \author{I.~Be\v{s}li\'c \inst{\ref{bonn},\ref{paris}} \orcid{0000-0003-0583-7363}
    \and 
        A.~T.~Barnes \inst{\ref{bonn}, \ref{eso}} \orcid{0000-0003-0410-4504}
    \and 
        F.~Bigiel \inst{\ref{bonn}} 
    \and 
        M.~J.~Jim\'enez-Donaire \inst{\ref{oan}, \ref{yebes}}
    \and
        A.~Usero \inst{\ref{oan}}
    \and 
        J.~D.~Henshaw \inst{\ref{liverpool}}
    \and 
        C.~Faesi \inst{\ref{connecticut}}
    \and
        A.~K.~Leroy \inst{\ref{osu}}
    \and
        E.~Rosolowsky \inst{\ref{alberta}}
    \and 
        J.~S.~den Brok \inst{\ref{bonn}, \ref{cfa}} \orcid{0000-0002-8760-6157}
    \and 
        M.~Chevance \inst{\ref{zah}} \orcid{0000-0002-5635-5180}
    \and
        C.~Eibensteiner \inst{\ref{bonn}} \orcid{0000-0002-1185-2810}
    \and
        K.~Grasha \inst{\ref{australia1}, \ref{australia2}}
    \and 
        R.~S.~Klessen \inst{\ref{ita}, \ref{iwr}} \orcid{0000-0002-0560-3172}
    \and
        J.~M.~D.~Kruijssen \inst{\ref{munich}} \orcid{0000-0002-8804-0212}
    \and
        D.~Liu \inst{\ref{eso}}
    \and
        S.~Meidt \inst{\ref{belgium}}
    \and 
        J.~Neumann \inst{\ref{mpia}} \orcid{0000-0002-3289-8914}
    \and   
        L.~Neumann  \inst{\ref{bonn}}
    \and
        H.~Pan \inst{\ref{taiwan}} \orcid{0000-0002-1370-6964}
    \and
        J.~Puschnig \inst{\ref{bonn}}
    \and
        M.~Querejeta \inst{\ref{oan}}
    \and
        E.~Schinnerer \inst{\ref{mpia}}
    \and 
        T.~G.~Williams \inst{\ref{oxford}} \orcid{0000-0002-0012-2142}
   }
      \institute{ 
            Argelander-Institut f\"ur Astronomie, Universit\"at Bonn, Auf dem H\"ugel 71, 53121 Bonn, Germany \label{bonn} 
        \and
            LERMA, Observatoire de Paris, PSL Research University, CNRS, Sorbonne Universit\'es, 75014 Paris, France \label{paris}
        \and
            European Southern Observatory (ESO), Karl-Schwarzschild-Stra{\ss}e 2, 85748 Garching, Germany \label{eso} 
        \and    
            Observatorio Astronómico Nacional (IGN), C/Alfonso XII, 3, E-28014 Madrid, Spain \label{oan}          
        \and 
            Centro de Desarrollos Tecnológicos, Observatorio de Yebes (IGN), 19141 Yebes, Guadalajara, Spain \label{yebes}
        \and
            Astrophysics Research Institute, Liverpool John Moores University, 146 Brownlow Hill, Liverpool L3 5RF, UK \label{liverpool}
        \and
            Department of Physics, University of Connecticut, Storrs, CT, 06269, USA \label{connecticut}
        \and 
            Department of Astronomy, The Ohio State University, 140 West 18th Ave, Columbus, OH 43210, USA \label{osu}
        \and 
            Department of Physics, University of Alberta, Edmonton, AB T6G 2E1, Canada \label{alberta}
        \and 
            Center for Astrophysics, Harvard \& Smithsonian, 60 Garden St., 02138 Cambridge, MA, USA \label{cfa}
        \and 
            Universit\"{a}t Heidelberg, Zentrum f\"{u}r Astronomie, Institut f\"{u}r Theoretische Astrophysik, Albert-Ueberle-Str.\ 2, 69120 Heidelberg, Germany \label{zah}
        \and 
            Research School of Astronomy and Astrophysics, Australian National University, Canberra, ACT 2611, Australia \label{australia1}
        \and 
            ARC Centre of Excellence for All Sky Astrophysics in 3 Dimensions (ASTRO 3D), Australia \label{australia2}
        \and 
            Universit\"{a}t Heidelberg, Zentrum f\"{u}r Astronomie, Institut  f\"{u}r Theoretische Astrophysik, Albert-Ueberle-Strasse 2, 69120 Heidelberg, Germany \label{ita}
         \and
            Universit\"{a}t Heidelberg, Interdisziplin\"ares  Zentrum f\"{u}r Wissenschaftliches Rechnen, Im Neuenheimer Feld 225, 69120 Heidelberg, Germany \label{iwr}
        \and 
            Technical University of Munich, School of Engineering and Design, Department of Aerospace and Geodesy, Chair of Remote Sensing Technology, Arcisstr. 21, 80333 Munich, Germany \label{munich}
        \and 
            Sterrenkundig Observatorium, Universiteit Gent, Krijgslaan 281 S9, B-9000 Gent, Belgium \label{belgium}
         \and
            Max Planck Institute for Astronomy, K\"onigstuhl 17, D-69117 Heidelberg, Germany \label{mpia}
        \and 
            Department of Physics, Tamkang University, No.151, Yingzhuan Road, Tamsui District, New Taipei City 251301, Taiwan \label{taiwan}
        \and
            Sub-department of Astrophysics, Department of Physics, University of Oxford, Keble Road, Oxford OX1 3RH, UK \label{oxford}
            \\
             }        

   \date{Received 26 07, 2023; accepted 29 02, 2024}

  \abstract
{Investigating molecular gas tracers, such as hydrogen cyanide (HCN), to probe higher densities than CO emission across nearby galaxies remains a challenge. This is due to the large observing times required to detect HCN at high sensitivity and spatial resolution. Although $\sim$kpc scale of HCN maps are available for tens of galaxies, higher-resolution maps are still scarce.}
{We aim to study the properties of molecular gas, the contrast in intensity between two tracers that probe different density regimes (the HCN\Jone/CO\Jtwo\, ratio), and their kinematics across NGC~253, one of the closest starburst galaxies. With its advanced capabilities, the Atacama Large Millimeter/submillimeter Array (ALMA) can map these features at high resolution across large field of view and uncover the nature of such dense gas in extragalactic systems.} 
   {We present new ALMA Atacama Compact Array and Total Power (ACA+TP) observations of the HCN emission across NGC~253. The observations cover the inner 8.6$'$ of the galaxy disk at a spatial resolution of 300 pc. Our study examines the distribution and kinematics of the HCN-traced gas and its relationship with the bulk molecular gas traced by CO\Jtwo. We analyze the integrated intensity and mean velocity of HCN and CO along each line of sight. We also use the SCOUSE software to perform spectral decomposition, which considers each velocity component separately.}
   {We find that the denser molecular gas traced by HCN piles up in a ring-like structure at a radius of 2\,kpc. The HCN emission is enhanced by 2 orders of magnitude in the central 2 kpc regions, beyond which its intensity decreases with increasing galactocentric distance. The number of components in the HCN spectra shows a robust environmental dependence, with multiple velocity features across the center and bar. We have identified an increase in the HCN/CO ratio in these regions, which corresponds to a velocity component that is likely associated with a molecular outflow. 
   We have also discovered that the ratio between the total infrared luminosity and dense gas mass, which is an indicator of the star formation efficiency of dense gas, is anticorrelated with the molecular gas surface density up to approximately 200 M$_{\odot}$\,pc$^{-2}$. However, beyond this point, the ratio starts to increase. }
   {We argue that using information about spectroscopic features of molecular emission is an important aspect of understanding molecular properties in galaxies. }
   

   \maketitle
%

\section{Introduction} 
\label{sec:introduction}

The densest structures of molecular clouds are found to be sites of star formation across galaxies \citep{gao_2004a,lada_2012,longmore_2014}. However, since the primary constituent of molecular gas, H$_2$ does not probe the coldest, densest parts of these clouds, other molecular lines are needed to observe star-forming gas and be used to constrain various properties. In the past, researchers have primarily used CO, the second most abundant molecule in the universe, to observe molecular clouds \citep[see review][]{bolatto_2013b}. CO emits relatively bright signals within the millimeter and (sub)millimeter regime, and its abundance scales with H$_2$. The assumption that the CO emission traces the overall molecular gas content of the interstellar medium (ISM) is commonly used within the literature \citep[e.g.][]{bigiel_2008,leroy_2008,tacconi_2010,schruba_2011,cormier_2014,genzel_2015,saintonge_2017}. 

In most cases, the low-J CO lines can be used to estimate the total H$_2$ mass, although they might not be reliable bulk tracers in extreme environments with, e.g., high cosmic-ray ionization rates \citep{bisbas_2015} or with intense UV radiation and/or low-metallicity \citep{pak_1998}.

CO is a good tracer of the cloud-scale surface density \citep{sun_2018}, but it does not reveal information about the star-forming part of molecular clouds. Therefore, to probe such dense regions ($n>10^3\,\mathrm{cm^{-3}}$, $200\,$M$_{\odot}$pc$^{-2}$), astronomers observe molecular lines with high critical densities \citep[high-critical density molecules {-}][]{shirley_2015}, such as those of HCN, HCO$^+$, N$_2$H$^+$, defined in this paper as "dense gas tracers". However, observing this gas at extragalactic distances is challenging; many preferred molecular lines commonly used in Galactic studies \citep{forbrich_2014,pety_2017,kauffmann_2017}, such as N$_2$H$^+$, can only be easily detected towards the centers of bright, nearby galaxies NGC~253 \citep{martin_2021} and NGC~6946 \citep{eibensteiner_2022, jimenez-donaire_2023}. 

The $J=1\xrightarrow{}0$ transition of the hydrogen cyanide, HCN, is one of the brightest high-critical density molecular lines commonly studied within the extragalactic literature \citep[e.g.][]{gao_2004a, usero_2015, gallagher_2018a, jimenez19, beslic_2021, sanchez-garcia_2022}. The comparison between low and high-critical density lines, such as CO\Jone\ and HCN\Jone, yields an approximate gauge of the intensity contrast between two tracers probing different density regimes, as the latter requires significantly higher densities for collisional excitation compared to the low-J CO lines \citep[$n_{\rm crit, eff}$\,$>$\,$10^4$~cm$^{-3}$ versus $>$\,$10^2$~cm$^{-3}$ \mbox{-} ][]{shirley_2015}. This contrast offers the best currently available observational constraint on changes in the underlying density distribution in other galaxies \citep[e.g., \,][]{leroy_2017a}.

From the observational point of view, HCN surveys in extragalactic systems found a tight and linear correlation between HCN luminosity and star formation rate \citep[SFR, e.g.,][]{gao_2004a, jimenez19}. This correlation is approximately linear in logarithmic space, spanning more than ten orders of magnitude, and covers a wide range of physical scales: from dense clumps and cores (a few pc - e.g., \citep[a few pc - e.g.][]{wu_2010} within the Milky Way, to galaxy disks and galaxy centers \citep[e.g.][]{gao_2004a,gracia_carpio_2008}. Although it has been argued that the linear relation between HCN and SFR suggests that gas above a specific density threshold starts forming stars \citep{lada_2012}, systemic variations in the IR/HCN ratio imply that not all dense gas is equally efficient. For example, the Central Molecular Zone (CMZ), known to be the inner 500\,pc region of our Galaxy, shows an order of magnitude lower SFR than those predicted from measurements of dense molecular gas \citep{longmore_2013a, henshaw_2022}, implying a possibility of the non-universality of the star formation efficiency \citep{maclow_2004, padoan_2011, federrath_2012, kruijssen_2014a, semenov_2015}. This environmental impact on the dense gas ability for star formation has also been observed across other galaxies. While most dense gas mass is found in centers of galaxies, but this gas is less efficient at star formation than the gas in the rest of the disk \citep{gallagher_2018a, jimenez19}. Observed variations in IR/HCN could be caused by non-steady effects dominant at small spatial scales such as stellar feedback and galactic shear, which become averaged out on larger scales.

In order to fully understand star formation process, it is important to connect what we learn from studying extragalactic sources with studies of individual star-forming regions within our own Milky Way. To do this, we need to observe molecular clouds in extragalactic samples at a high sensitivity and on small spatial scales, while also covering a broad range of different environments and physical conditions. This is a crucial step towards connecting studies of the Milky Way, where we can examine the substructure of individual molecular clouds \citep[e.g.,][]{pety_2017}, with studies of more distant galaxies that contain vastly different environments, such as starburst galaxies \citep{garcia_burillo_2012}. To achieve this, it is essential to observe nearby galaxies that span the range of scales needed to benchmark our understanding of local clouds and high-z galaxies. Due to their proximity, we can map dense molecular gas content across nearby galaxies at high spatial resolution and sensitivity. Moreover, the high molecular surface brightness of these sources, known as \sigmamol, can provide a sensitive mapping of the HCN emission that we can now observe using advanced observing facilities such as ALMA. Recent studies of HCN at higher resolutions, approaching molecular cloud-scales of around 100\,pc, have been focused mainly on the brightest regions of nearby galaxies, such as in M~51 \citep{querejeta_2019}, a larger part of the disk of NGC~3627 \citep{beslic_2021}, the center of NGC~6946 \citep{eibensteiner_2022}, and the inner ring of NGC~1068 \citep{sanchez-garcia_2022}.

\begin{figure*}[t!]
	\includegraphics[width=\textwidth]{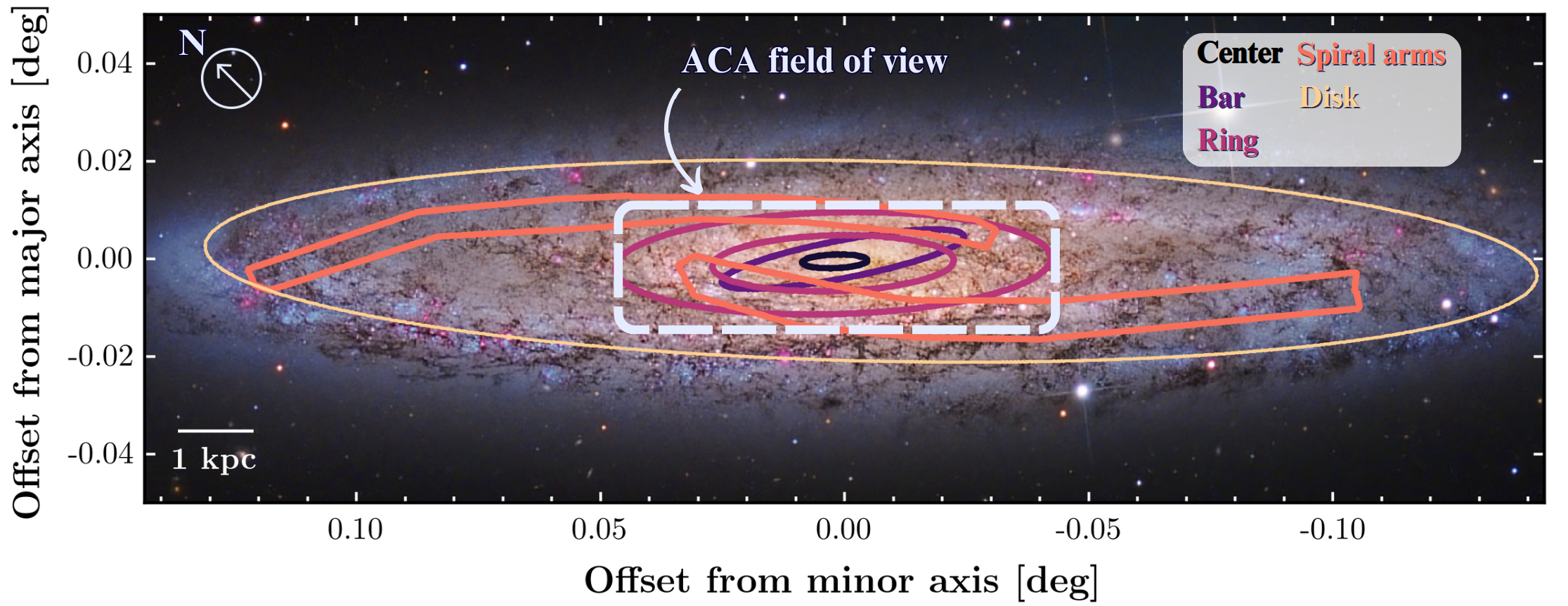}
	\includegraphics[width=\textwidth]{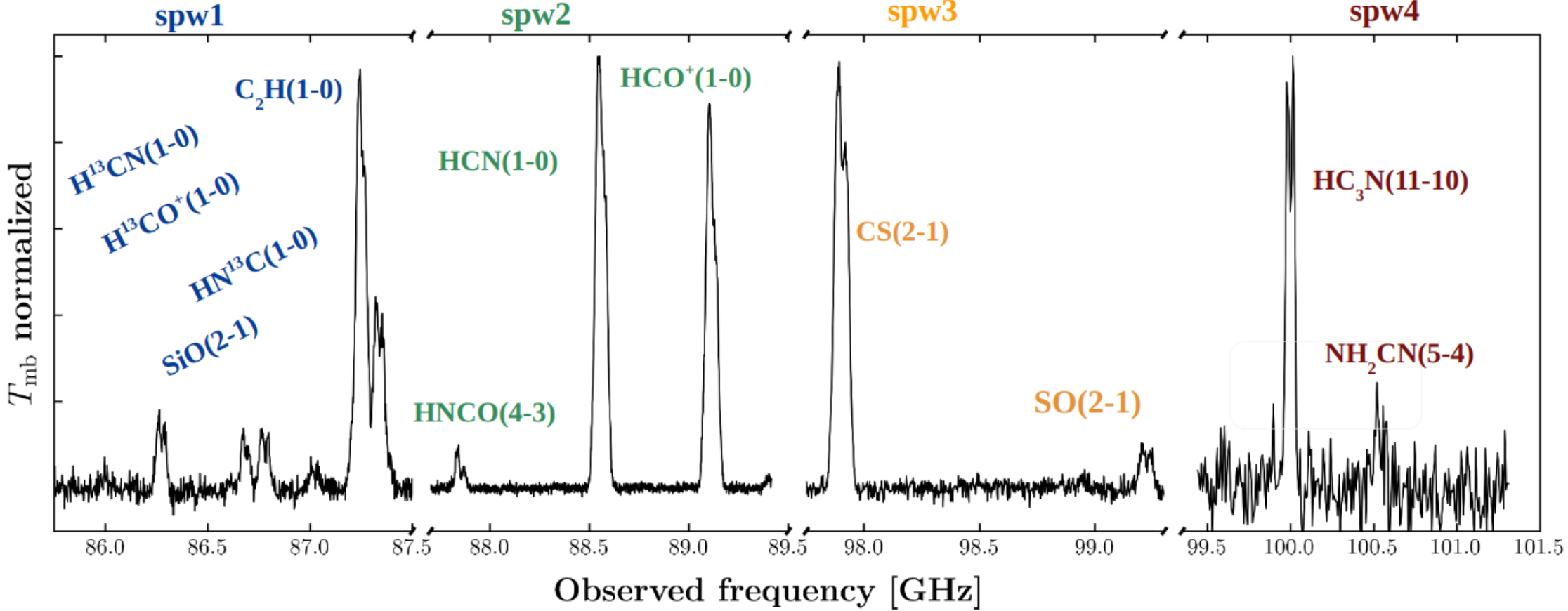}
    \caption{\textbf{Top panel:} NGC~253 composite (RGB and H$\alpha$) image (credits: Terry Robinson). The overlaid gray dashed rectangle shows the total $u-\upsilon$ coverage consisting of 25 pointing mosaic obtained by the ACA 7-meter for this work. We indicate dynamical environments present in the disk of NGC~253: the center, bar, ring, and spiral arms \citep{querejeta_2021}. \textbf{Bottom panel:} An observed spectrum taken towards the center of NGC~253. Our observations cover frequencies from 85.7 to 101.5 GHz. The term spectral window i (i takes values from 1 to 4) is abbreviated as spw$_\mathrm{i}$, and we mark all detected molecular emission lines.}
    \label{fig:ngc253}
\end{figure*}

In this paper, we answer some of the science questions related to high-critical density molecular emission, such as the following:
\begin{enumerate}
\item How is dense molecular gas distributed across the disk of NGC~253 (Section\,\ref{sec:line_of_sight})?
\item What is the kinematics of the HCN-tracing gas, and how does it compare with the CO emission (Section\,\ref{sec:scouse})?
\item What is the role of dense gas in star formation across different environments found in NGC~253 (Section\,\ref{sec:star_formation})?
\end{enumerate}
To address these questions, we use ALMA observations of the closest starburst galaxy outside the Local Group, the Sculptor galaxy, NGC~253. The main properties of this galaxy are listed in Table\,\ref{tab:NGC0253_prop}. NGC~253 (see its composite image in Figure\,\ref{fig:ngc253}) is the highly inclined, starburst galaxy \citep{rieke_1980} located in the southern hemisphere. Due to its proximity \citep[1 arcsecond corresponds to 17 pc at a distance of 3.7\,Mpc {-}][]{anand_2020}, NGC~253 represents an ideal target for high-resolution studies to understand the nature of its kpc nuclear region \citep{bolatto_2013, leroy_2015, walter_2017, holdship_2021} of which the center appears to be undergoing an intense phase of active star formation. The star formation rate within the center is SFR=2\,M$_\mathrm{\odot}$yr$^{-1}$ \citep{leroy_2015}, which is 50\,$\%$ of the SFR found in the whole disk of this galaxy \citep{sanders_2003}.

The central one kpc of NGC~253 shows rich molecular emission (see our detected molecular lines in the bottom panel of Figure\,\ref{fig:ngc253}), demonstrated in \citet{martin_2006}, \citet{aladro_2011}, and \citet{Meier_2015}, who detected 50 molecular species at 3\,mm wavelength (e.g.\ C$_2$H$_5$OH, HOCN, HC$_3$H). In addition, ALMA Comprehensive High-resolution Extragalactic Molecular Inventory (ALCHEMI) mapped a continuous frequency range from (sub)mm to mm wavelengths in the nuclear region of NGC~253 \citep{martin_2021}. NGC~253 has a plethora of ancillary data covering a variety of gas phases, from hot \citep{lopez_2022}, atomic \citep[e.g.][]{heckman_2000}, molecular \citep{houghton_1997, mauersberger_1996, paglione_2004}, and the most recent ACA observations of CO\Jtwo\ which we use in this work \citep{faesi_2023}, to ionized gas \citep{arnaboldi_1995}. 

\begin{table}
\centering
\caption{Main properties of NGC~253.}
\begingroup
\setlength{\tabcolsep}{8pt} 
\renewcommand{\arraystretch}{1.2} 
\begin{tabular}{cc}
    \hline \hline
    Property & Value \\
    \hline
    Name & NGC~0253 (Sculptor galaxy) \\
    Hubble type $^{(a)}$ & SABb \\
    Center RA (J2000) & 00h47m33s \\
    Center DEC (J2000) & -25d17m19s  \\
    Inclination, $i$ $\mathrm{[^{\circ}]}$ $^{(a)}$ & 76$\pm$6 \\ 
    Position angle, $PA$ $\mathrm{[^{\circ}]}$ $^{(a)}$ & 52 \\ 
    Distance, $D$ $\mathrm{[Mpc]}$ $^{(b)}$ & 3.7 \\
    $r_{25}$ [\arcmin] $^{(b,c)}$ & 5.12 \\
    $V_\mathrm{sys, hel}$ [km\,s$^{-1}$] $^{(d)}$ & 258 \\
    SFR [M$_{\odot}$ yr$^{-1}$] $^{(e)}$ & 4.2 \\
    log$_{10}(M_*$) [$\mathrm{M}_\odot$] $^{(f)}$ & 10.5\\
    \hline\hline
\end{tabular}
\endgroup
\begin{minipage}{0.95\columnwidth}
    \vspace{1mm}
    (a) 
    \cite{hlavacek-larrondo_2011}. \\
    (b) Distance from \citet{anand_2020}.\\
    (c) Radius of the $B$-band 25th magnitude isophote.\\
    (d) Systemic velocity from \citet{casasola_2011}. \\
    (e) Star formation rate calculated from the IR luminosity \citep{sanders_2003}. \\
    (f) Integrated stellar mass based on 3.6\,\textmu m\ emission, taken from the PHANGS-ALMA survey paper \citep{leroy_2021b}.\\
\end{minipage}
\label{tab:NGC0253_prop}
\end{table} 

Although NGC~253 has been well-studied over the last few decades, most investigations have focused on its inner 1-2\,kpc nuclear region. The molecular gas outside the galaxy center needs to be better understood, particularly the properties of its dense phase and its kinematics on larger scales in a wide range of disk environments. 

In this paper, we present new ACA observations across the disk of the NGC~253. These data have a spatial resolution of 300\,pc, covering a large part of the disk of this galaxy ($8\times3$\arcmin$^2$) and probe different dynamical features in this galaxy. The paper is structured as follows. Section \ref{sec:obs} describes observations, data reduction, additional data sets used throughout this work, and present moment maps of the HCN and CO\Jtwo. Next, we present our results: Section\,\ref{sec:line_of_sight} describes the distribution of the line of sight HCN emission and the measured HCN/CO\Jtwo\ ratio. Section\,\ref{sec:scouse} presents the decomposed line of sight HCN emission and measured properties of such decomposed emission and describes results on measured velocity dispersion. Moreover, we answer Question\,3 in Section\,\ref{sec:star_formation}. Our findings are discussed in Section\,\ref{sec:discussion}. Finally, we summarize and outline our most important results in Section\,\ref{sec:summary}.

\section{Observations and data reduction}
\label{sec:obs}
\begin{figure*}[t!]
	\includegraphics[width=\textwidth]{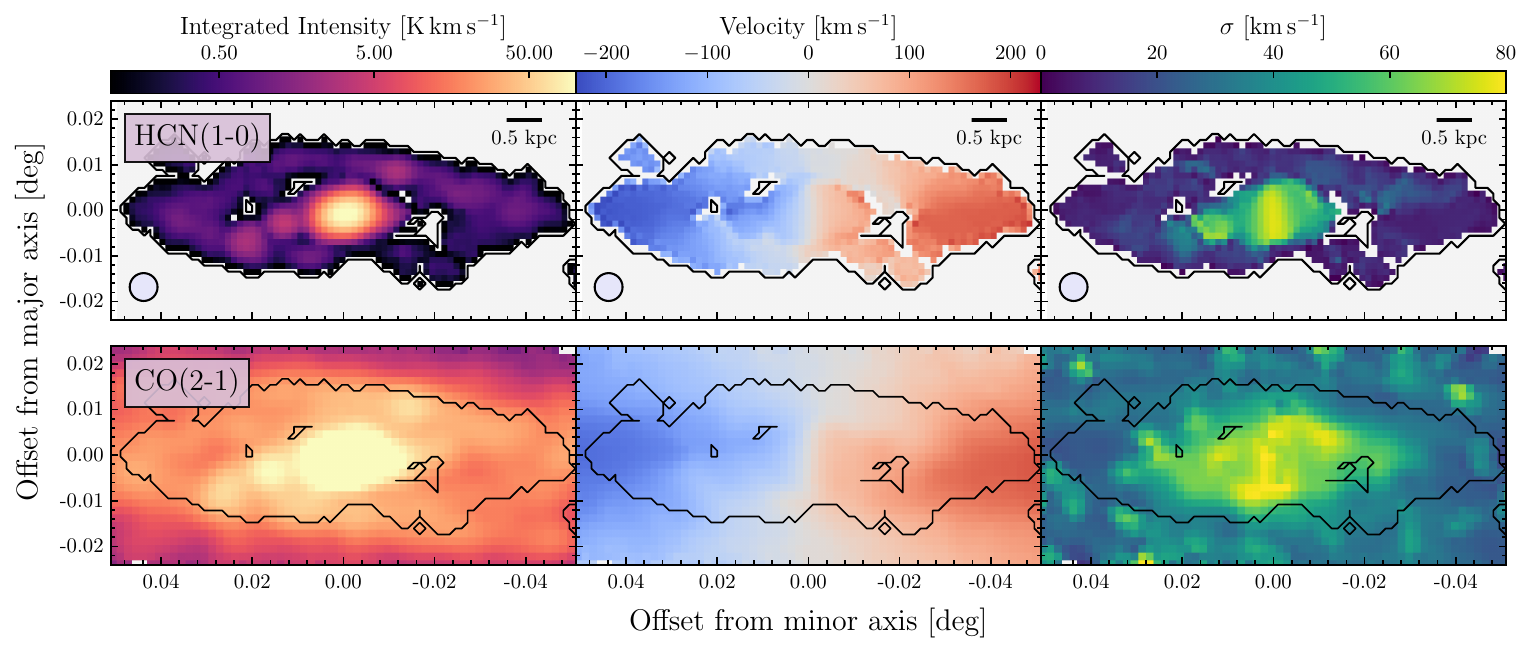}
    \caption{\textbf{Top row:} HCN\Jone\ moment maps: integrated intensity map (left), centroid velocity map (middle), and velocity dispersion (right). The beam size of 22\,\arcsec is shown in the left corner of each panel. \textbf{Bottom row:} CO\Jtwo\ moment maps, in the same order as for HCN\Jone. CO\Jtwo\, is convolved to a beam size of 22\,\arcsec, and regrided to match the grid of HCN data. The black line shows the outline of the HCN emission from the top panels. We fix the colorscale for each moment map to highlight the differences and similarities in HCN and CO\Jtwo\ emission. All maps are rotated so that the x- and y-axis show the angular distance from the NGC~253's minor and major axis (see Table\,\ref{tab:NGC0253_prop}).}
    \label{fig:moment_maps}
\end{figure*}

\subsection{ALMA+ACA observations}

The observations (project ID: 2019.2.00236.S) of NGC~253 presented in this paper are part of the ALMA Cycle 7 and were obtained in December 2019 and January 2020. We observe molecular line emission using the Atacama Compact Array (7m - ACA), which included the Total Power (TP) array in Band 3 (84 – 116 GHz), covering several ''typical'' extragalactic tracers of high-density gas, shown on the bottom panel in Figure\,\ref{fig:ngc253}. The total observing time is 22.9 and 44.1 hours for the 7\ meter array and the TP antennas, respectively. The observed field of view contains 25 7m pointings with the primary beam full-width half maximum (FWHM) of 57.6\,\arcsec, that covers a $114$\arcsec$\times516$\arcsec ($\sim$2$\times$9 kpc$^2$) region of the inner disk of NGC~253 (see Figure \ref{fig:ngc253}). The beam size is $21.40$\arcsec$\times10.36\,$\arcsec, and the beam position angle is 83.5 degrees. From our ancillary CO\Jtwo\, observations (Section\,\ref{sec:smol}) and SFR images (Section\,\ref{sec:ssfr}), we find that the region mapped within the ACA comprises 90$\%$ of the total molecular mass of NGC~253 and 85$\%$ of its total SFR. Thus, our observations cover the brightest regions in molecular gas and the most star-forming parts in NGC~253. 

Our interferometric observations are sensitive to the emission from angular scales of $\sim$14\,\arcsec ($\sim$240\,pc) to the largest angular scale obtained by the TP of $\sim$100\,\arcsec ($\sim$1.7\,kpc). The obtained sensitivity is $\sim$2.5\,mK (7.6\,mJy/beam) at the angular and spectral resolution of 22\,\arcsec and 10\,km\,s$^{-1}$. The spectral setup consisted of 4 spectral windows with $\sim1.8$\,GHz width and 0.4\,MHz resolution laid out as at the bottom of Figure\,\ref{fig:ngc253}. We detected 13 molecular lines, but here, we focus on the HCN\Jone\ line covered by spectral window 2, labeled as spw2 on Figure\,\ref{fig:ngc253}. 

The raw data calibration is done in Common Astronomy Software Applications (CASA {-} \citealp{mcmullin_2007}), version 6.5, and the imaging and post-processing, including short-spacing correction using the PHANGS-ALMA processing pipeline \citep{leroy_2021a}. Firstly, we flag all the emission lines within each band to calculate and subtract the continuum emission. The 7m data was imaged using CASA's standard \textsc{tclean} procedure. We use the CO\Jtwo-based clean mask \citep{leroy_2021a, leroy_2021b} as the initial clean mask for our imaging, which is then combined with the single-scale clean during the imaging. The HCN 7m flux within the cleaning mask is 99$\%$ of the total flux within the full image. 

The single-dish data are also processed using the PHANGS-ALMA total power processing pipeline \citep{herrera_2020}, included in the PHANGS-ALMA pipeline \citep{leroy_2021a}. In the final data reduction step, we combine 7m observations with TP for the missing short-scale emission using the standard CASA task \textit{feather}. The total flux from the interferometric data is $70\%$ of the total flux measured from the single-dish data. The final data cube is additionally primary beam corrected and convolved to have a circular beam size of 22\,\arcsec ($\sim$370\,pc) and channel width of 10\,\kms.

In addition, we compute the dense gas mass, $M_\mathrm{dense}$, assuming a constant $X_\mathrm{HCN} = 10\,[\mathrm{M_{\odot}\,(\mathrm{K\,km\,s^{-1}\,pc^2})^{-1}] }$ conversion factor \citep{gao_2004a}:

\begin{equation}
    M_\mathrm{dense} [\mathrm{M_{\odot}}] = X_\mathrm{HCN}\cdot L_\mathrm{HCN} \cdot \cos{i},
\label{eq:mdense}
\end{equation}

\noindent where $i$ is the galaxy inclination reported in Table\,\ref{tab:NGC0253_prop}.

\subsection{Environmental masks}
\label{sec:env}

To separate regions with different characteristics in NGC~253, we use environmental masks defined for this galaxy based on infrared data \citep{querejeta_2021}. They have four regions: the disk, ring, bar, and center. Additionally, NGC~253 has two spiral arm features, visible in optical data \citep{pence_1980} and constrained from the 3.6 and 4.5\,\textmu m observations obtained by the Spitzer telescope, as part of the S$^4$G survey \citep{munoz-mateos_2015, herrera-endoqui_2015}. Therefore, we also define a spiral arm mask in this work as follows. We use the unsharp-masked near-infrared Herschel PACS $70$\textmu m data to locate spiral features, and then we fit those in polar ($\rho,\theta$) space as linear functions. The width of such constructed spiral arms is defined manually in the SAO-NASA ds9 software \citep{ds9}. Finally, we add these spiral arm regions to the existing environmental mask, shown in Figure\,\ref{fig:ngc253}. 
We distinguish five environments in NGC~253: the nuclear region lies within the inner $\sim$0.5\,kpc region, and the bar feature is located in the inner two \,kpc region. The ring is at radii between 2 and 5\,kpc, and the dust lanes are at $r_\mathrm{gal}=5$\,kpc. All remaining pixels not assigned to any of these environments belong to the disk.

\subsection{ALMA-CO\Jtwo\ observations from PHANGS}
\label{sec:smol}

In this work, we use ancillary ALMA ACA+TP CO\Jtwo\ observations (PI: C.~Faesi, 2018.1.01321.S), included in the PHANGS-ALMA survey \citep[PI: E.~Schinnerer - ][]{leroy_2021b,faesi_2023}. The angular resolution of these data is 8\,\arcsec, and the channel width is 2.5\,km\,s$^{-1}$. In the final step, we convolve the CO\Jtwo\ data cube to match the working angular and spectral resolution of 22 \arcsec and 10\,\kms\ and regrid to a common pixel scale. We compute the molecular surface density using the following equation:

\begin{equation}
  \Sigma_\mathrm{mol} [\mathrm{M_{\odot} pc^{-2} }] = \alpha_\mathrm{CO}\cdot \dfrac{I_\mathrm{CO(2-1)}}{R_\mathrm{21}} \cdot \cos{i} = \alpha_\mathrm{CO}\cdot \dfrac{I_\mathrm{CO(2-1)}}{\Sigma_{\mathrm{SFR}}^{0.15}} \cdot \cos{i} ,
\label{eq:sigma_mol}
\end{equation}

\noindent where $\alpha_\mathrm{CO}$ is the metallicity-dependent conversion factor taken from \cite{sun_2020b}, $I_\mathrm{CO(2-1)}$ is the CO\Jtwo\, integrated intensity, $R_{21}$ is the CO\Jtwo/CO\Jone\, line ratio \citep[e.g.,][]{sandstrom_2013, denbrok_2021, yajima_2021, leroy_2022}, and $i$ is the same as in Equation\,\ref{eq:mdense}. In this work, we use the scaling relation: $R_{21}\propto \Sigma_\mathrm{SFR}^{0.15}$ \citep[see][and references therein]{sun_2023}. The $\alpha_\mathrm{CO}$ has a range from 4.2 (at the centre) to 20 $[$M$_{\odot}/(\mathrm{K\,km\,s^{-1}\,pc^2})]$ (at the outskirts). The validity of the $\alpha_\mathrm{CO}$ factor, depends on the mass scales probed. For masses greater than $10^5$\,M$_{\odot}$, the statistical assumptions underlying the $\alpha_\mathrm{CO}$ factor remain valid \citep[e.g.][]{dickman_1986}, which is achieved in our work. We present the $\Sigma_\mathrm{mol}$ map in Figure\,\ref{fig:smol_map} in Appendix\,\ref{sec:app-smol}.

\subsection{Star formation rate}
\label{sec:ssfr} 

To estimate the star formation rate surface density ($\Sigma_\mathrm{SFR}$), we use a combination of IR Herschel data from the KINGFISH (Key Insights on Nearby Galaxies: a Far-Infrared Survey with Herschel) survey \citep{kennicutt_2011}. To compare our results with recent similar studies \citep[e.g.][]{jimenez19}, we choose to calculate $\Sigma_\mathrm{SFR}$ using Herschel bands at $\lambda=$70, 160, and 250 microns. First, we convolve our Herschel maps to a final resolution of 22\,\arcsec\, using the kernels defined in \citet{aniano_2011} and match the coordinate grid with the final HCN data image. We calculate the total infrared surface density ($\Sigma_\mathrm{TIR}$) following the prescription from \citep{galametz_2013}:

\begin{equation}
    \Sigma_\mathrm{TIR} [\mathrm{W\,kpc^{-2}}] = \sum_{j} c_{j}\cdot \Sigma_{j} [\mathrm{W\,kpc^{-2}}],
\label{eq:tir}
\end{equation}
where $c_{j}$ is the coefficient and $\Sigma_{j}$ is the surface density at a band j, listed in Table\,\ref{tab:coeff_tir} in Appendix\,\ref{sec:app-sfr}. We then calculate $\Sigma_\mathrm{SFR}$ from $\Sigma_\mathrm{TIR}$ and correct for the galaxy inclination $i$ \citep{galametz_2013}:
\begin{equation}
    \Sigma_\mathrm{SFR} [\mathrm{M_{\odot}\,yr^{-1}\,kpc^{-2}}] = 1.48\cdot10^{-10}\, \Sigma_\mathrm{TIR} [\mathrm{L_{\odot}\,kpc^{-2}}] \cos{i}.
\end{equation}

\noindent We show the $\Sigma_\mathrm{SFR}$ map in Figure\,\ref{fig:sfr_map} in Appendix\,\ref{sec:app-sfr}.

\subsection{Moment maps}
\label{sec:res_mom_maps}

We show moment maps, i.e., the line of sight integrated intensity (moment 0) map, centroid velocity map (moment 1), and velocity dispersion map (moment 2) in Figure\,\ref{fig:moment_maps}, at 22\,\arcsec resolution (370 pc) for HCN\Jone\ (top row) and CO\Jtwo\ data (bottom row).

The integrated intensity is computed as a sum of the emission along each line of sight (i.e., we integrate along the velocity axis). We first create a CO\Jtwo\ based mask. This mask is produced by selecting all voxels with a signal-to-noise ratio higher than 4 and then expanding it to include all contiguous voxels with signal-to-noise $>2\sigma$ level. Next, we apply this mask to the HCN and CO\Jtwo\ data cubes. The moment maps are derived from the masked data cubes using the \texttt{python} package SpectralCube \citep{spectral_cube}.

The HCN emission shows a clumpy structure and compact emission, prominent in the center of the galaxy and also concentrated along the ring and the bar. There are spots of bright HCN emission located at the contact points of the bar, dust lanes, and ring, which we also see in the $\Sigma_\mathrm{SFR}$ (Figure\,\ref{fig:sfr_map} in Appendix\,\ref{sec:app-sfr}). These bright spots of HCN emission are expected to be found in the regions where the bar and ring intersect, as these are interfaces where gas orbits converge \citep{kenney_1994,beuther_2017}. Moreover, we see CO\Jtwo\ bright spots co-spatial with those seen in HCN. 

The moment one map is the intensity-weighted mean velocity map. We show these in the middle column of Figure\,\ref{fig:moment_maps} for both HCN and CO\Jtwo. The velocity map is corrected for the systemic velocity (Table\,\ref{tab:NGC0253_prop}). The outskirts of the map (left and right side) show the highest velocity difference, i.e., along the major axis (around $-230$ and $+230$ \kms\ on the eastern and western parts, respectively). The northeastern side of the galaxy is blue-shifted. HCN and CO\Jtwo\ velocities look mainly consistent across the disk of NGC~253 (we discuss this in detail in Section\,\ref{sec:scouse}). We expect their consistency from the assumption that HCN traces denser gas than the CO, thus probing denser molecular substructures traced by the CO emission. We note a few regions within which we observe discrepancy in centroid velocities of HCN and CO\Jtwo\, and further discuss when considering spectral complexity along the line of sight in Section\,\ref{sec:scouse}.

The second moment map represents the velocity dispersion of the observed emission. The broadest line profiles are observed towards the center of this galaxy in both HCN and CO emission. Apart from the center, we also note other regions with relatively large velocity dispersion ($\sim50\,$km/s) in the HCN and CO\Jtwo\ emission, such as the bar and partially the ring. CO\Jtwo\ shows higher velocity dispersion than the HCN, and this difference becomes more notable at the outer parts of our map.

\section{Line of sight HCN emission}
\label{sec:line_of_sight}
\subsection{Radial distribution of the line of sight HCN intensity}
\label{sec:hcn_radial_profile}

\begin{figure}[t!]
	\includegraphics[width=\columnwidth]{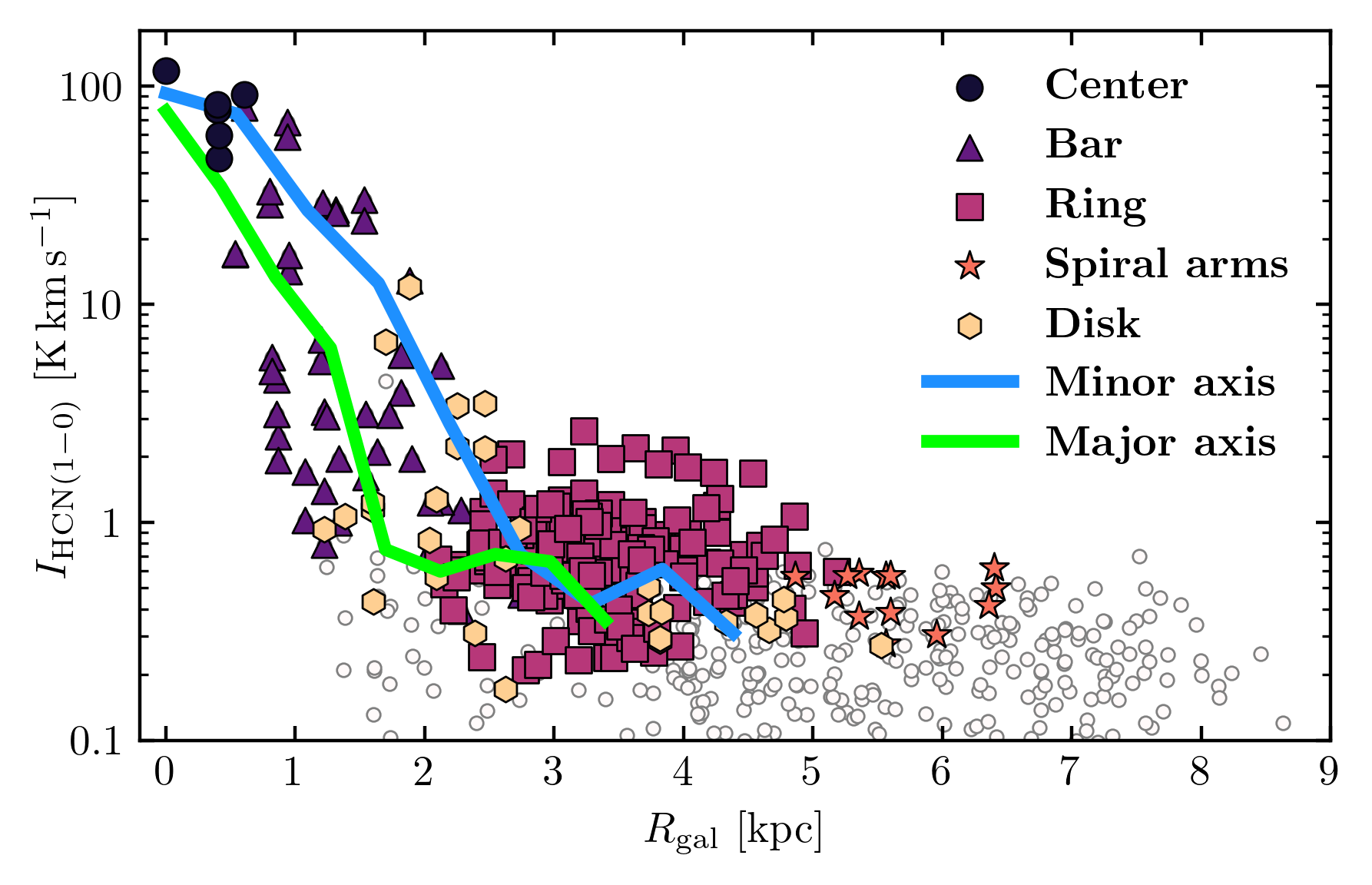}
    \caption{HCN integrated intensity as a function of galactocentric radius. The coloured, filled points represent pixels with a signal-to-noise ratio of 3 and above, whereas open points have a signal-to-noise ratio lower than 3.}
    \label{fig:hcn_radial_profile}
\end{figure}

\begin{figure*}[h!]
	\includegraphics[width=\textwidth]{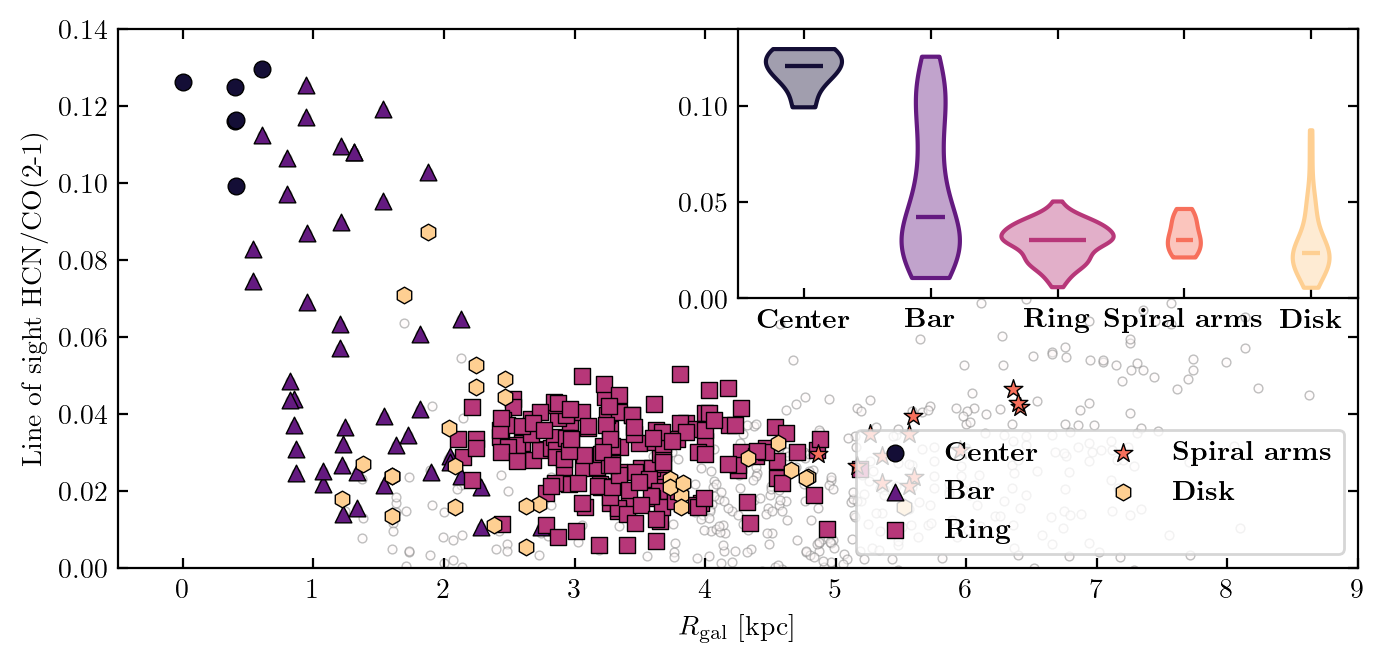}
    \caption{Radial distribution of the HCN/CO\Jtwo\, intensity ratio across NGC~253. Sight lines are colour-coded based on the environment. We show points with signal-to-noise ratio above 3 in intensity and the peak brightness temperature. A small panel in this figure shows distribution of HCN/CO\Jtwo\ line intensity ratio per environment in NGC~253: the center, bar, ring, spiral arms and disk. The mean line ratio measured in each environment is shown as a horizontal line.}
    \label{fig:hcn_co21}
\end{figure*}

To investigate the HCN emission across NGC~253, we resample the HCN map on a hexagonal grid where adjacent pixels are half-beam spaced. We show the radial distribution of the HCN integrated intensity in Figure\,\ref{fig:hcn_radial_profile}. We detect significant HCN emission up to $\sim6.5$\,kpc. The high inclination of this galaxy causes large deprojected distances from the center along the minor axis (i.e., \ the minor axis effect). After inspection of the deprojected distances along the major axis, where we do not expect to encounter such impact, we conclude that the farthest data point from the center of the map has a deprojected distance of 5.3\,kpc, assuming an axisymmetric distribution.
Consequently, data points at radii higher than the 5.3\,kpc should be taken cautiously when interpreting such radial trends. In addition, this effect will blur the observing trends. The uncertainty of the inclination (Table\,\ref{tab:NGC0253_prop}) affects the calculated radial distances by a factor of 0.4.

Next, we extract sight lines from each environment by applying the environmental mask \citep{querejeta_2021} and color-code the points by the environment, following the color scheme used in Figure\,\ref{fig:ngc253}. The HCN intensity distribution in NGC~253 follows the CO\Jone\ distribution described in \cite{sorai_2000, paglione_2004}, i.e.\ the "twin-peaks" morphology, typical for barred galaxies \citep{kenney_1992}. The brightest HCN emission is located at the center of the galaxy, which is also seen in other galaxies: at kpc resolution in the EMPIRE \citep{jimenez19} and ALMOND \citep{neumann_2023}, in M~51 \citep{bigiel_2016}, in nearby galaxies at similar spatial scales \citep[$\sim$0.5\,kpc,][]{gallagher_2018a}, and at higher spatial resolution in NGC~3627 \citep{beslic_2021}. There is a steep decrease in HCN intensity by two orders of magnitude along the bar. Moreover, the HCN intensity varies by order of magnitude at distances of $r_\mathrm{gal}>2$\,kpc, coincident with the ring and spiral arms. With increasing distance from the center of NGC~253, we also note a decrease in the HCN emission, particularly in the inner 1-2\,kpc region within the bar where HCN intensity decreases steadily by order of magnitude, similar to the barred galaxy NGC~3627 \citep{beslic_2021}.


Ancillary observations of the CO\Jone\ emission at similar spatial resolution (16$''$ $\sim$250\,pc) across NGC~253 \citep{sorai_2000, paglione_2004} show similar radial trends. \cite{sorai_2000} observed a secondary peak in CO\Jone\ surface density around $r_\mathrm{gal}$=2\,kpc, located at the end of the bar. In our case, we observe a similar increase in the CO\Jtwo\ emission at a distance between $3-4$\,kpc from the center of NGC~253, right outside the bar. 
In addition, we observe local enhancements in HCN in regions where the bar and the ring overlap and a decrease in HCN emission in the outermost parts of our map. By comparison, \cite{sorai_2000} found a decrease in CO\Jtwo\ surface brightness by two orders of magnitude in the inner 1-2\,kpc region of NGC~253, and the intensity after the second peak observed at 2\,kpc is steadily decreasing. Moreover, the rotation curve of NGC~253 derived from the CO\Jone\ emission \citep{koribalski_1995, sorai_2000} flattens in the ring, which suggests that molecular gas at these positions starts losing its angular momentum and infalling towards the center \citep{sorai_2000}.

\subsection{Line of sight HCN/CO\Jtwo\ intensity ratio}
\label{sec:los_hcn-co}

\begin{figure*}[t!]
	\includegraphics[width=\textwidth, keepaspectratio=true]{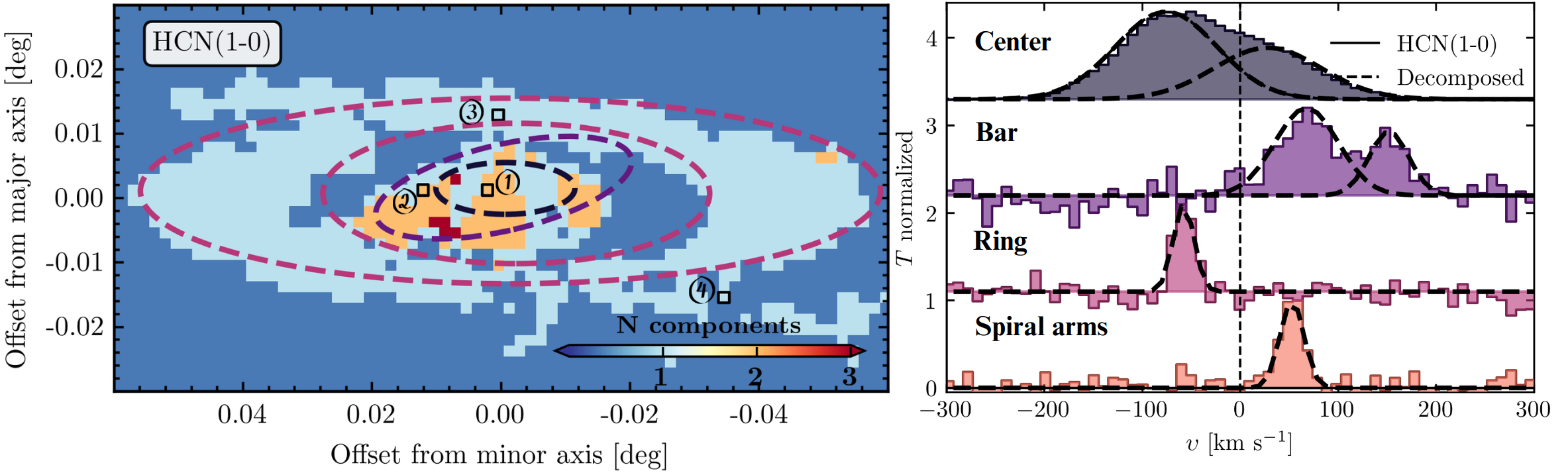}
    \caption{\textbf{Left} Map of NGC~253 showing number of velocity components in HCN emission across each pixel derived from SCOUSE. \textbf{Right:} Example of HCN spectra from each environment shown in Figure\,\ref{fig:ngc253} and from coloured points on the left panel. The black dashed lines show each Gaussian velocity component derived from the SCOUSE spectra decomposition analysis \citep{henshaw_2016a, henshaw_2019}. }
    \label{fig:scouse1}
\end{figure*}

The ratio of HCN to CO\Jone\ is often used within the literature to constrain the dense gas fraction, \fdense\ \citep{leroy_2017a}. Knowing the scaling between $J=\Jone$ and $J=\Jtwo$ intensity ratio \citep[e.g.,][]{sandstrom_2013, zschaechner_2018, denbrok_2021,leroy_2022}, it is possible to use HCN/CO\Jtwo\ for determining the \fdense. In the case of NGC~253, we are particularly interested in seeing how this line ratio varies across the galaxy and search for any environmental dependence. 

We calculate the HCN/CO\Jtwo\ integrated intensity ratio from the hexagonally, half-beam-sized sampled grid data points at 22\,\arcsec. The uncertainty of the HCN/CO\Jtwo\ ratio is computed from the uncertainties of their integrated intensities, using the standard error propagation technique \citep[see e.g.][]{beslic_2021,eibensteiner_2022}. 
Figure \ref{fig:hcn_co21} shows the HCN/CO\Jtwo\ line intensity ratio as a function of a distance from the center of NGC~253. The radial distribution of HCN/CO\Jtwo\ follows the trends seen in the HCN emission (Figure\,\ref{fig:hcn_radial_profile}). It decreases towards larger galactocentric distances. The highest values (from 0.1 to 0.13) of this line ratio are found within the center of NGC~253. The transition from the central region to the disk is sharp, especially in the inner 2\,kpc region of NGC~253 along the bar, where the HCN/CO\Jtwo\ decreases by an order of magnitude. This finding is similar to the results reported in previous studies \citep{bigiel_2016, jimenez19, gallagher_2018a, querejeta_2019}, whereas the weak radial variation of HCN/CO\Jtwo\ intensity ratio across NCG~3627 was found in \citet{beslic_2021}. In the rest of the environments, particularly in the ring, spiral arms and the disk, the HCN/CO\Jtwo\ intensity ratio does not vary significantly (values between 0.01 and 0.05). We note exceptionally high values of the HCN/CO\Jtwo\ intensity ratio at distances larger than 6\,kpc, which arises from the fact that at these distances CO\Jtwo\ emission appears to decrease more rapidly than the HCN.

To quantify HCN/CO\Jtwo\ in these environments, we show the distribution of this line ratio in each region in the form of a violin plot on the upper right corner of Figure \ref{fig:hcn_co21}. In this panel, the length of each violin corresponds to the range of HCN/CO\Jtwo\, intensity ratios. In contrast, the horizontal line indicates the mean measured in each environment. Points located in the center of NGC~253 have the highest mean of HCN/CO\Jtwo\ of 0.1197$\pm$0.0002, which is about 4 times higher than the median values found in the rest of the galaxy. The bar shows the widest distribution and possible bimodal behavior. The first group of sight lines has a high HCN/CO\Jtwo\ ratio of 0.1 and above, close to those found in the center. The second group of points contains HCN/CO\Jtwo\ values similar to those seen in the ring and the rest of the regions (from 0.01 to 0.08). Therefore, the bar represents an intermediate region that bridges the center and the rest of the environments found in NGC~253. The enhancement in the HCN/CO\Jtwo\ ratio found has a few possible implications, which we discuss in more detail in Section\,\ref{sec:outflow}. 

The observed HCN/CO\Jtwo\ line ratio shows agreement within the literature data for other galaxies. Similar to \citet{gallagher_2018a}, \citet{jimenez19}, and \citet{beslic_2021}, this line ratio has the highest values in the inner $\sim$500\,pc region of the galaxy, i.e.\ the vast majority of the dense molecular gas relative to bulk molecular content is found within the center of a galaxy. \citet{knudsen_2007} found HCN/CO\Jone\ ratio in the inner 1\,kpc region to be 0.08. By applying the $R_{21}$ line ratio of 0.8 \citep{zschaechner_2018}, we obtain higher values of HCN/CO\Jone. As opposed to HCN/CO\Jtwo\ in the barred galaxy NGC~3627 \citep{beslic_2021}, we see a stronger environmental dependence of HCN/CO\Jtwo\ and a significant decrease toward higher galactocentric radii. 

\section{Decomposed HCN emission}
\label{sec:scouse}
\begin{figure*}[t!]
	\includegraphics[width=\textwidth]{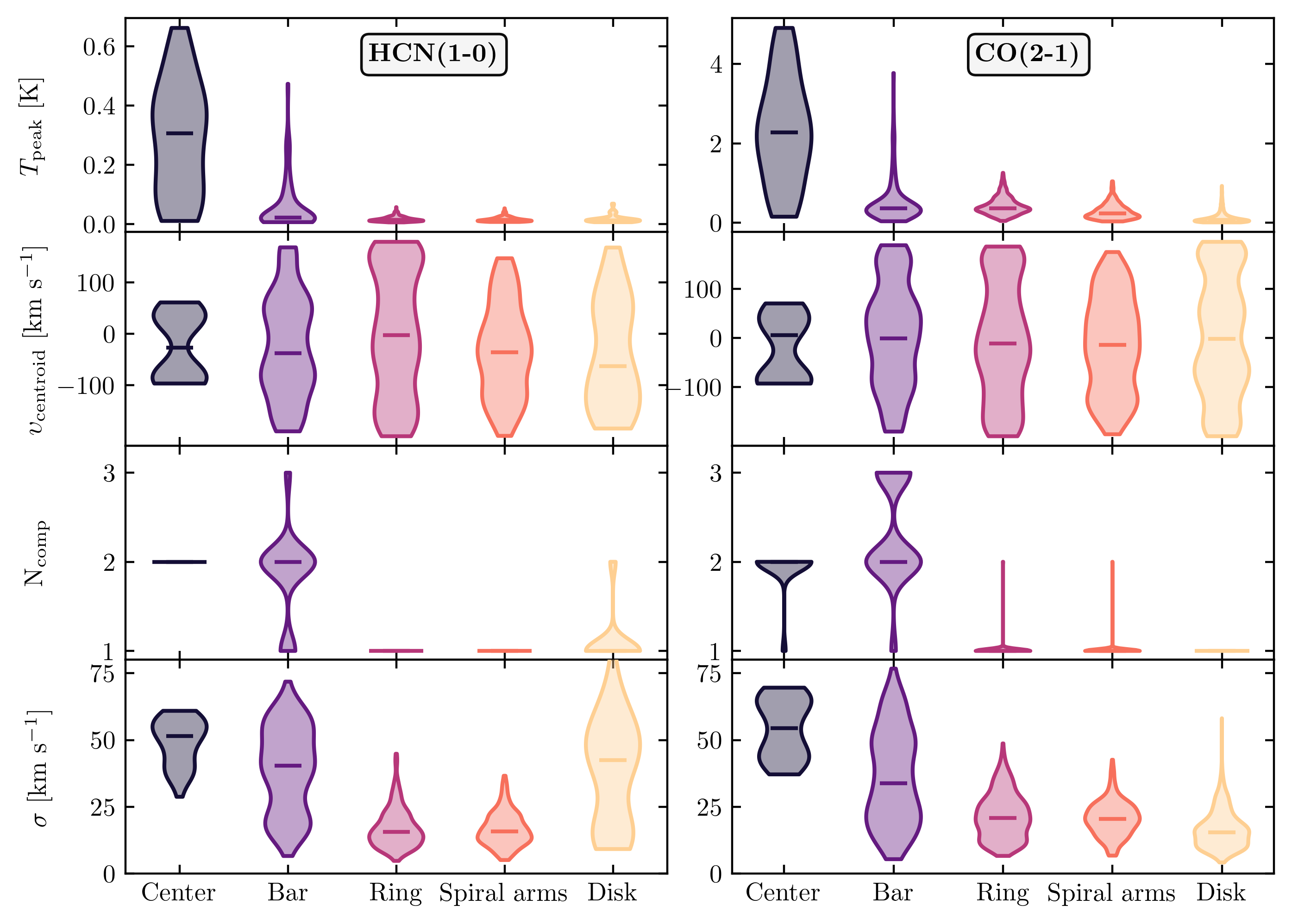}
	\caption{The Gaussian fit parameters: peak brightness temperature (first row), centroid velocity (second row), number of components (third row), and velocity dispersion (bottom row) derived from SCOUSE. Left column shows results from decomposing the HCN emission, whereas the right column shows output parameters computed for the CO\Jtwo\ emission. We show results for each environment. Distribution for HCN emission along the disk is represented as a circle since there was no converging fits.}
    \label{fig:scouse2}
\end{figure*}

To characterize the gas structure along each line of sight and its variation with the galactic environment, we study the spectra of the HCN emission in 5 different dynamical zones. We use a Semi-automated multi-COmponent Universal Spectral-line fitting Engine \citep[SCOUSE][]{henshaw_2016a, henshaw_2019} to decompose the observed emission into separate components along each line of sight. This program is developed to analyze spectral lines and extract information (e.g., centroid velocity, line width) from modeling the shape of a spectral line. Appendix\,\ref{sec:app-scouse} provides a detailed description of using SCOUSE. 
We show the number of HCN velocity components identified by SCOUSE on the left panel of Figure\,\ref{fig:scouse1}. Figure\,\ref{fig:scouse1} shows examples of 4 spectra originating from different dynamical regions (center, bar, ring, and spiral arms {-} each marked with a black rectangle on the left panel of Figure\,\ref{fig:scouse1}).
In particular, lines of sight from the central region and the bar have multiple line profiles, whereas spectra towards the ring structure, spiral arms, and the disk are mainly narrow and single-peaked (the right panel in Figure\,\ref{fig:scouse1}). Similarly, \citet{sorai_2000} found compound velocity structures in CO\Jone, particularly within the bar. However, this is not due to the line broadening caused by galactic rotation within the beam. Instead, \citet{sorai_2000} proposed that strong non-circular motions cause the existence of two velocity components along the line of sight.

In Figure\,\ref{fig:scouse2}, we present our results from applying the SCOUSE analysis to HCN and CO\Jtwo\ datasets. Each panel of the figure shows the distribution of fitting parameters describing a Gaussian line shape derived from SCOUSE, which include the number of components, peak temperature, centroid velocity, and velocity dispersion. Furthermore, these parameters are divided according to each environment.

As shown in Figure\,\ref{fig:scouse1}, each environment contains different spectral features in HCN and CO\Jtwo. CO\Jtwo\ profiles are generally brighter than HCN, and the spectra within the center, followed by those in the bar, have the highest temperature peaks. The distribution of $T_\mathrm{peak}$ within the center has the largest dynamical range compared to the $T_\mathrm{peak}$ determined in spectra from other environments. 

The third row in Figure\,\ref{fig:scouse2} summarizes what has been demonstrated in Figure\,\ref{fig:scouse1}. The highest number of velocity components is seen towards the bar and nuclear region. HCN spectra are predominantly double and triple-peaked in these regions, unlike HCN spectra in the ring, spiral arms, and disk. We find similar spectral complexity of the CO\Jtwo\ emission. In addition, we identify several pixels whose CO\Jtwo\ spectrum shows three velocity components in the bar and only a few pixels containing three HCN velocity components. Two scenarios could explain the lower frequency of multiple components observed in the HCN emission than in the CO\Jtwo\ spectrum. The first explanation is that the HCN emission is generally fainter than CO\Jtwo. In this scenario, assuming that each CO\Jtwo\ has its own HCN partner and the new CO\Jtwo\, component along the same line of sight has a considerably lower signal-to-noise ratio than the previous one, it would explain why more HCN components were not detected. In the other case, since CO\Jtwo\ and HCN trace overall different gas densities, it may occur that certain CO\Jtwo\ clouds do not contain denser subregions possibly traced by HCN. The number of CO\Jtwo\ velocity components towards each sight line within the ring and spiral arms is mostly one, except for a few lines of sight where we identify two CO\Jtwo\ velocity features within the same pixel, possibly due to higher CO\Jtwo\ sensitivity.

\begin{figure*}[t!]
    \centering
    \includegraphics[width=\textwidth]{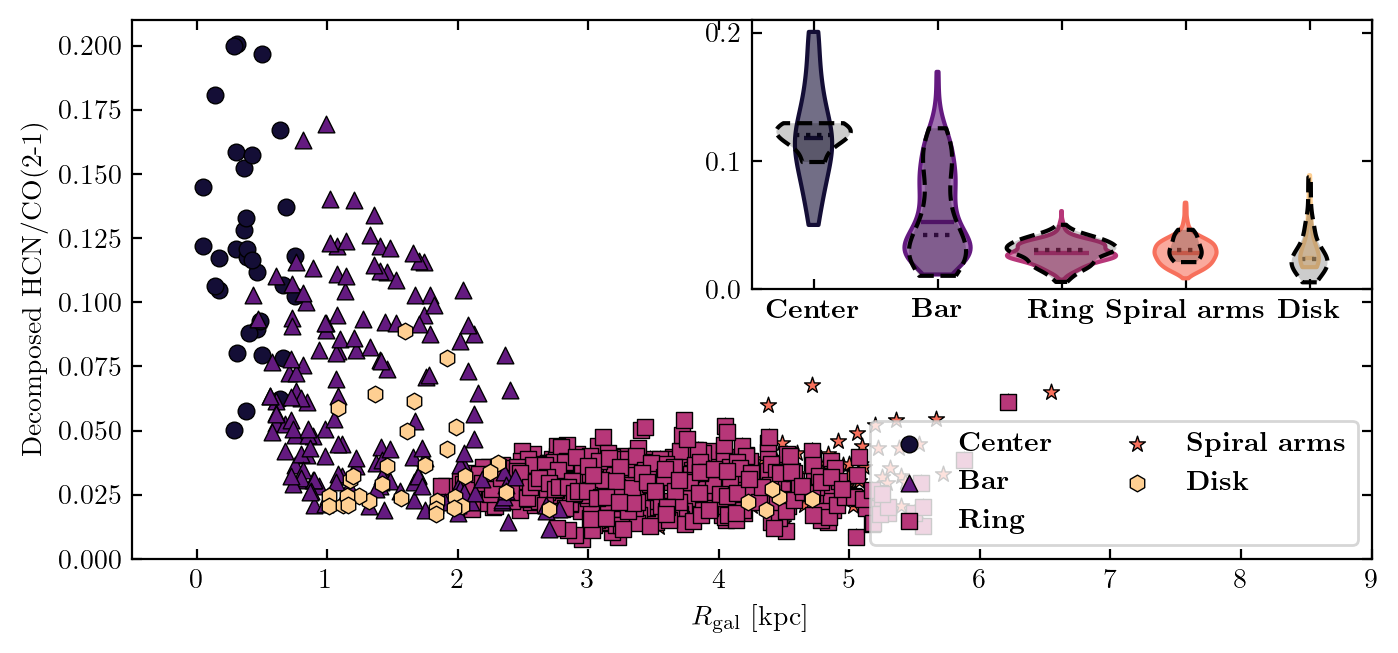}
    \caption{Same as in Figure\,\ref{fig:hcn_co21}, but the HCN/CO\Jtwo\ intensity ratio is calculated using SCOUSE.}
    \label{fig:scouse_hcn_co21_ratio}
\end{figure*}

Next, we comment on the individual velocity dispersion of HCN and CO\Jtwo. Here, the environmental dependence becomes prominent as in $T_\mathrm{peak}$. Central sight lines show the broadest lines, followed by sight lines from the bar. The distribution of HCN and CO\Jtwo\ velocity dispersion in the center has a bimodal behavior. The broadest is the $\sigma$ distribution within the bar in HCN and CO\Jtwo\. The distribution of $\sigma$ in HCN and CO\Jtwo\, is compact in the rest of the environments. The highest median in $\sigma$ is found within the center (55\,km\,s$^{-1}$ and 65\,km\,s$^{-1}$ for the HCN and CO\Jtwo). Interestingly, within the bar, we find higher median values in HCN (40\,km\,s$^{-1}$) than in CO\Jtwo\, emission (30\,km\,s$^{-1}$). In the ring and spiral arms, the median $\sigma$ in CO\Jtwo\, is higher than in HCN. The lines of sight across the disk show broad HCN and CO\Jtwo\ components. These lines of sight are close to the center of NGC~253 (see Figure\,\ref{fig:hcn_radial_profile}).





\subsection{Decomposed HCN/CO\Jtwo\ line intensity ratio}
\label{sec:hcn_co21_scouse}

This section investigates the HCN/CO\Jtwo\ intensity ratio derived from SCOUSE and compares it with our previous findings described in Section\,\ref{sec:line_of_sight}. To directly compare HCN and CO\Jtwo\ intensities from each component derived from SCOUSE, it is necessary to match the detected velocity components in HCN with those found in the CO\Jtwo\ emission. As seen on the top left panel of Figure\,\ref{fig:scouse2}, the number of Gaussian components differs from region to region and between HCN and CO\Jtwo. This could result from the CO\Jtwo\, tracing more lower density gas at shifted velocity to which the HCN is not sensitive.

The first step in associating velocity components is to look at the centroid velocities determined for each component within each pixel in HCN and CO\Jtwo, particularly the difference in $\upsilon_\mathrm{centroid, HCN}$ and $\upsilon_\mathrm{centroid, CO\Jtwo}$. When the HCN velocity component is associated with the CO\Jtwo, their centroid velocities should be similar. Therefore, we consider that the HCN line is associated with the CO\Jtwo\ if:
\begin{equation}
\label{eq:centr}
    |\upsilon_\mathrm{centroid, HCN}-\upsilon_\mathrm{centroid, CO\Jtwo}|<\sigma_\mathrm{thresh}.
\end{equation}
We define the $\sigma_\mathrm{thresh}$ as following:
\begin{equation}
\label{eq:sigma_thresh}
    \sigma_\mathrm{thresh} = \dfrac{\sigma_\mathrm{CO(2-1)} + \sigma_\mathrm{HCN}}{2},
\end{equation}
and calculate this value for each pixel where we decompose the emission. Equations\,\ref{eq:centr} and \ref{eq:sigma_thresh} represent the overlap of the clouds. Such a criterion assures us that clouds with higher line widths have a less strict association criterion. After applying this threshold, we match 100\,$\%$ of HCN emission lines to those in CO\Jtwo\ in the center, ring, and spiral arms. We matched $78\%$ of HCN emission to the CO\Jtwo\ components within the bar. Around 60 percent of the CO\Jtwo\ flux entering this analysis (see Section\,\ref{sec:smol}) is associated with the HCN emission. In exceptional cases, when two components have similar centroid velocities, we take the one with higher amplitude, i.e., the $T_\mathrm{peak}$. 

Similarly, as in Figure\,\ref{fig:hcn_co21}, in Figure\,\ref{fig:scouse_hcn_co21_ratio} we show the component-by-component HCN/CO\Jtwo\ intensity ratio as a function of distance from the center of NGC~253 (left panel) and the distribution of this line ratio within each environment (right panel). The radial distribution of the component-by-component HCN/CO\Jtwo\ is similar to that of the line of sight HCN/CO\Jtwo\ (Figure\,\ref{fig:hcn_co21}). We also observe here the center showing the highest HCN/CO\Jtwo\ ratio, followed by the bar. Similarly, as in Figure\,\ref{fig:hcn_co21}, we observe a steady decrease in HCN/CO\Jtwo\ at distances of $1-2$\,kpc. By comparing the distributions of HCN/CO\Jtwo\ computed from SCOUSE with those presented in Section\,\ref{sec:los_hcn-co} (dashed violin shapes on the right panel in Figure\,\ref{fig:scouse_hcn_co21_ratio}), we see the biggest difference between these two approaches in the center. Although the means of these two distributions are similar, their shape is different, i.e., the decomposed HCN/CO\Jtwo\ distribution is significantly elongated than the line of sight distribution. The distributions of data points using both approaches within the bar are similar; they both show signs of bimodal behavior, but their mean values differ. The mean of the HCN/CO\Jtwo\ from SCOUSE is comparable to the mean values calculated from the line of sight line ratio. Moreover, distributions of HCN/CO\Jtwo\ of spiral arms and disk derived from SCOUSE are different than the line of sight HCN/CO\Jtwo, mainly because of a generally lower signal-to-noise ratio of the HCN emission in these environments, which resulted in a lower number of successfully decomposed lines of sight.


\subsection{HCN velocity dispersion}
\label{sec:sigma_hcn}

Molecular gas flows impact the dynamical state of the gas \citep{meidt_2018} and lead to collisions and gas crowding \citep[e.g.][]{beuther_2017}, thus possibly suppressing or enhancing star formation. This directly impacts the star formation efficiency, which is in line with turbulent models of star formation \citep{federrath_2013}. On the one hand, whether tied to galactic gas flows or star formation feedback, super-virial gas motions can reduce star formation efficiencies \citep{padoan_2011, meidt_2020a}. On the other hand, large velocity dispersions may signify elevated Mach numbers and a preferential build-up of dense gas \citep{krumholz_2005a, federrath_2012,jindra_2020}. 

The environments studied here represent a diversity of gas flows and conditions: strong shear in the center, elliptical streaming motions and strong radial inflows in the bar, a pile-up of gas in the ring, and strong spiral streaming motions. Here, we investigate the sensitivity of the gas velocity dispersion to these environments. We also explore how the velocity dispersion of dense gas compares to that of molecular gas. Dense gas arising from the small scales might be expected to be less sensitive to the flows present at or just beyond the cloud envelope, while its presence may reveal signatures of the critical role that colliding flows and shear have on the gas structure.

\begin{figure}[t!]
	\includegraphics[width=0.5\textwidth]{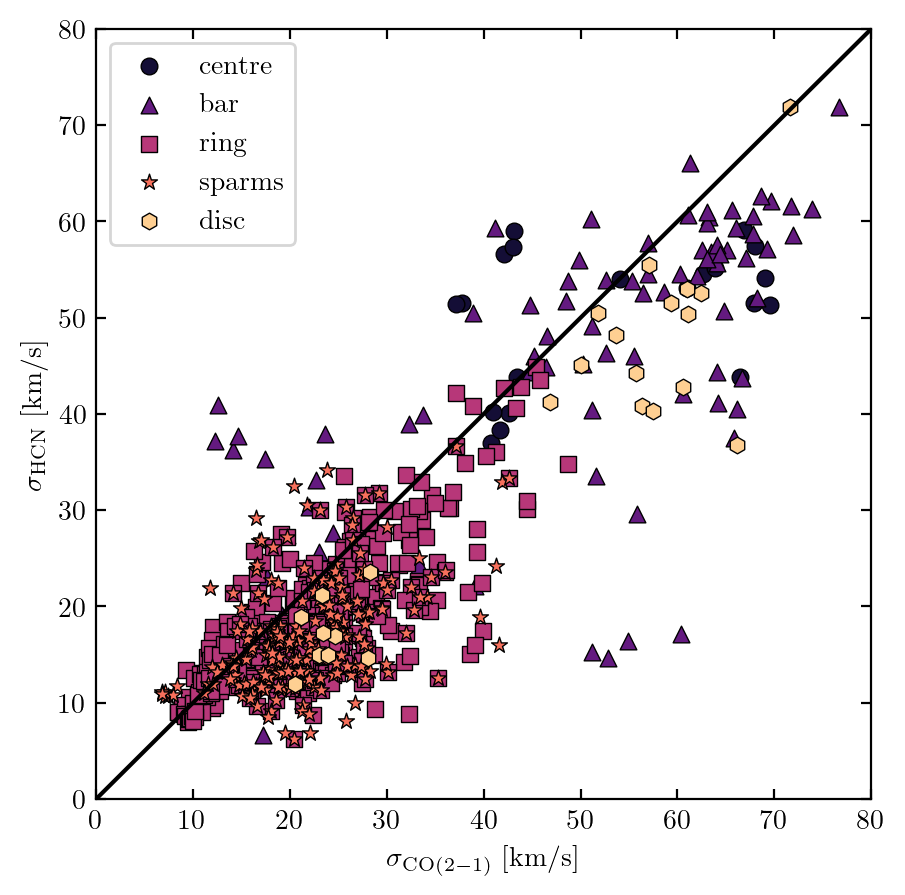}
    \caption{Component-by-component comparison between HCN (y-axis) and CO\Jtwo\ (x-axis) velocity dispersion derived from spectral decomposition using SCOUSE. We colour-code each point by the environment. The diagonal black solid line shows 1:1 ratio between x and y axis.}
    \label{fig:scouse3}
\end{figure}

We compare the observed velocity dispersion in NGC~253 for HCN and CO\Jtwo\ emission in Figure\,\ref{fig:scouse3}. The solid black line represents the one-to-one ratio in HCN and CO\Jtwo\ velocity dispersion. The HCN emission lines are narrower than the CO lines, as the data points typically lie below the diagonal line. This trend is particularly pronounced in the center and the bar, whereas data points are scattered around the solid black line at velocity dispersions below 50\,km\,s$^{-1}$. Central sight lines populate the higher-values region, whereas sight lines from the spiral arms (orange points) and the ring (pink points) have low values. Interestingly, bar sight lines are distributed over the full range of the observed values along both axes.

The general positive correlation in HCN and CO\Jtwo\, velocity dispersion in NGC~253 agrees with previous velocity dispersion measurements in M~51 \citep{querejeta_2019}, and across NGC~3627 \citep{beslic_2021}. Our result confirms expectations that higher-density gas (traced by the HCN in our case) has smaller turbulent velocity dispersion than lower-density gas (traced by the CO\Jtwo). This finding also agrees with the models that observed the building up of higher-density gas in the stagnation regions of convergent flows from the larger-scale turbulence. Consequently, the average velocity dispersion of the gas at these stagnation points will be smaller.

\begin{figure}
	\includegraphics[width=0.5\textwidth]{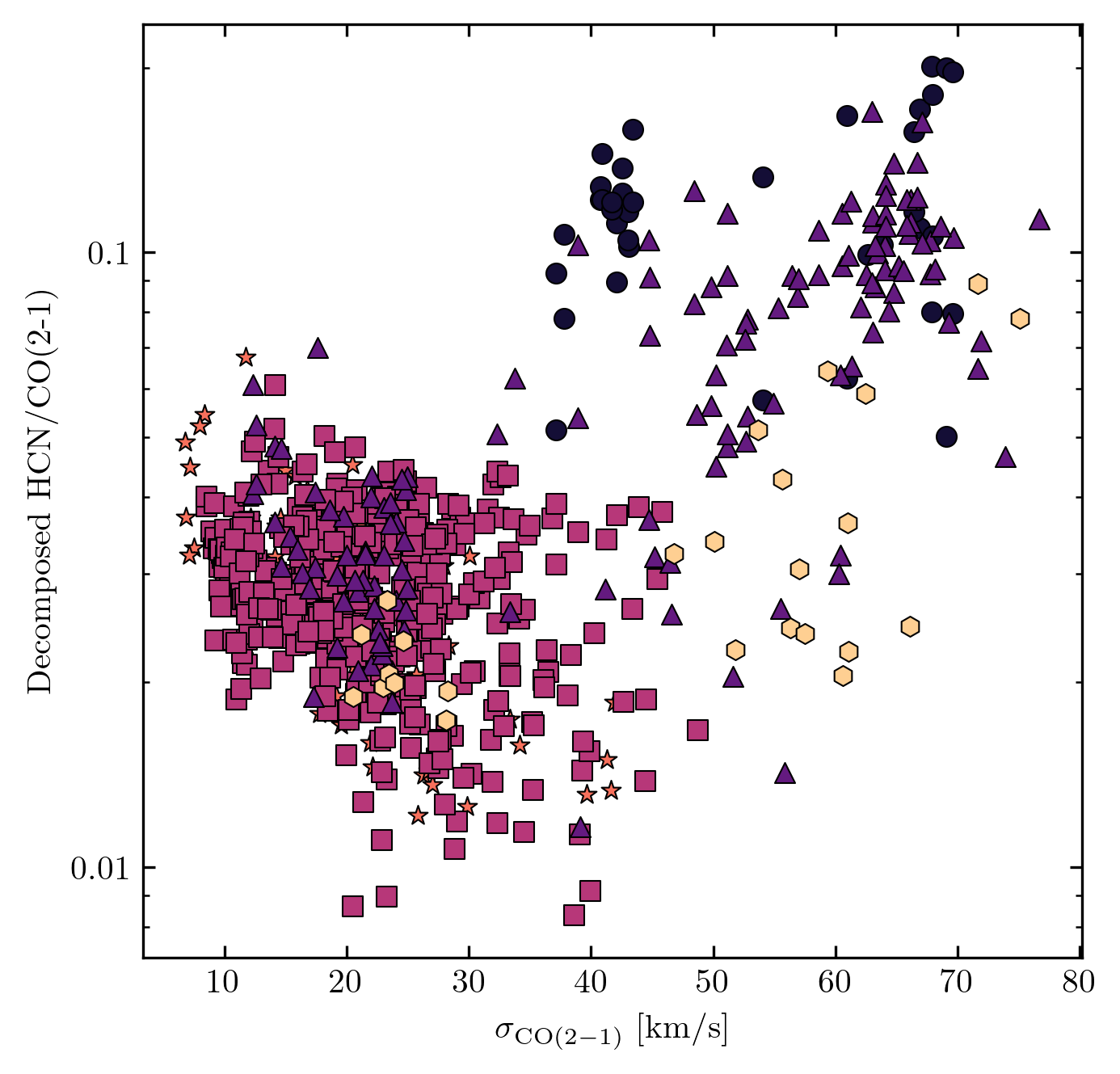}
    \caption{Decomposed HCN/CO\Jtwo\ intensity (y-axis) and the CO\Jtwo\ velocity dispersion (x-axis). We colour-code each point by the environment, similar as in Figure\,\ref{fig:scouse3}.}
    \label{fig:scouse4}
\end{figure}

\begin{figure}[t!]
	\includegraphics[width=0.5\textwidth]{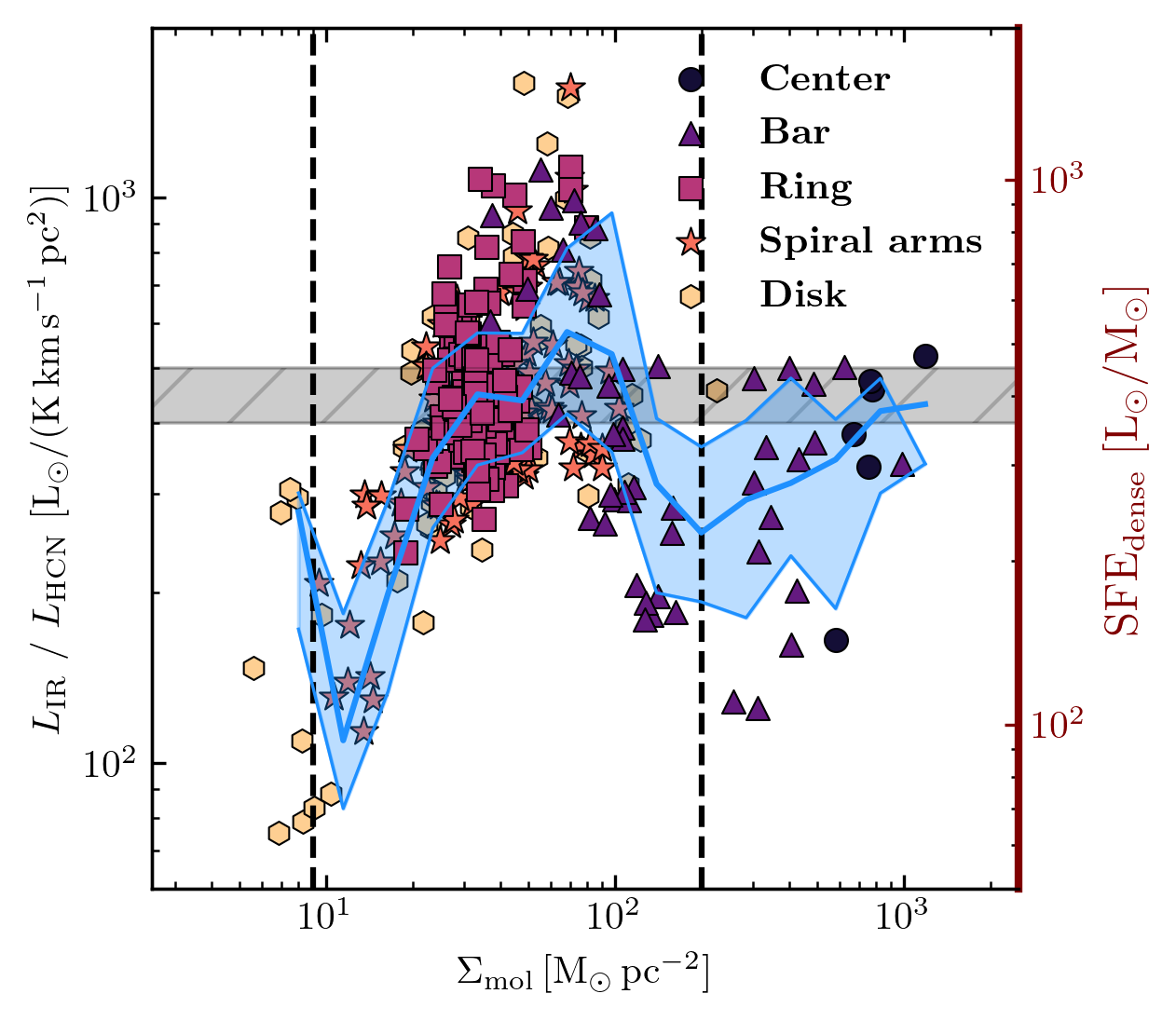}
    \caption{SFR/HCN luminosity ratio as a proxy for star formation efficiency of dense molecular gas as a function of $\Sigma_\mathrm{mol}$. Points are colour-coded according to their environments. Dashed vertical lines show the surface density thresholds of $\sim10\,\mathrm{M_{\odot}\,pc^{-2}}$ and $\sim200\,\mathrm{M_{\odot}\,pc^{-2}}$, previously defined in \citet{bigiel_2008}. Horizontal shaded region shows range of $400-500\,$L$_{\odot}$/M$_{\odot}$.}
    \label{fig:sfe_fdense}
\end{figure}

In Figure\,\ref{fig:scouse4}, we show decomposed HCN/CO\Jtwo\ intensity ratio derived from SCOUSE as a function of the CO\Jtwo\ velocity dispersion. Our data points populate two different parts of this figure. In particular, we note at $\sigma_\mathrm{CO\Jtwo}$ of $>45$\,km\,s$^{-1}$ higher HCN/CO\Jtwo\ values, which is a consequence of the environment, as these data points are located in the bar and the center. In the case of lower velocity dispersions, the 16\% and 84\% of observed HCN/CO\Jtwo\ lie in the range from $0.02$ to $0.04$. 

\section{Dense molecular gas and star formation}
\label{sec:star_formation}
\subsection{Environmental dependence of the star formation efficiency of the dense gas}
\label{sec:sfe_fdense} 

Previous extragalactic studies have shown different behavior of \sfedense\, across the disk of a galaxy. For example, regions with high stellar surface density and thus interstellar pressure, such as centers of galaxies, typically have relatively lower \sfedense, i.e., the ability of dense gas to form stars is significantly reduced \citep[e.g., ][]{longmore_2013, kruijssen_2014a, usero_2015, bigiel_2016, barnes_2017, jimenez19, beslic_2021, eibensteiner_2022}. In the following, we investigate properties of star formation efficiency of dense gas \sfedense\ (usually traced by the $L_\mathrm{IR}/M_\mathrm{dense}$) across NGC~253.

In Figure \ref{fig:sfe_fdense}, we show the TIR/HCN luminosity (TIR/$M_\mathrm{dense}$ on the right y-axis) ratio as a function of the molecular gas surface density. Dashed vertical lines in both panels show the gas surface density thresholds presented in \citet{bigiel_2008}. The left vertical line at $\sim10\,\mathrm{M_{\odot}\,pc^{-2}}$ indicates the surface density above which the molecular gas dominates over the atomic phase \citep{bigiel_2008}, whereas the right vertical line shows the surface density threshold of $\sim200\,\mathrm{M_{\odot}\,pc^{-2}}$, after which we enter the regime where dense molecular gas emission linearly scales with the star formation \citep[e.g.][]{gao_2004a, bigiel_2008}. The blue-shaded region corresponds to binned trends and their respective range of 16th and 84th percentiles. The horizontal shaded region shows a range of $400-500\,\mathrm{L_{\odot}/M_{\odot}}$.

Our HCN measurements are located in moderate density regimes (part of the bar, ring, spiral arms, and disk) and at high surface densities (lines of sight from the center and bar). That said, we do not have information about the HCN emission for molecular gas surface densities lower than a few $\mathrm{M_{\odot}\,pc^{-2}}$, where the atomic gas dominates the total gas surface densities \citep[e.g.][]{bigiel_2008}. The lowest values of \sfedense\ are measured in the disk of NGC~253. At the intermediate surface densities (9-200\,M$_\mathrm{\odot}$\,pc$^{-2}$), \sfedense\ rises with the cloud-scale molecular surface density up to $\sim100$M$_\mathrm{\odot}$\,pc$^{-2}$, after which it starts decreasing. The mean \sfedense\, increases up to $\sim500\,\mathrm{L_{\odot}/M_{\odot}}$ in the intermediate density regime, while the high surface density part shows nearly constant \sfedense\, of $\sim400\,\mathrm{L_{\odot}/M_{\odot}}$. The lowest \sfedense\ in this regime have lines of sight at the bar.

The center of NGC~253 is characterized by strong tidal forces and generally high pressure, typical also for other galaxies \citep[see, e.g.,][]{gallagher_2018b}. Therefore, tidal forces may impact the fraction of gas that is self-gravitating, which can lead to a reduction in the observed \sfedense\, when, at the same time, some part of the HCN-emitting gas is prevented from being in a state to collapse and form stars. Therefore, we suggest that the elevated average cloud density characteristic of the high-pressure central environment \citep{gallagher_2018b} implies that a greater portion of the cloud emits HCN in the central regions of NGC~253.

The observed behaviour of \sfedense\, presented on Figure\,\ref{fig:sfe_fdense} may result the non-constant conversion factor $X_\mathrm{HCN}$. When $X_\mathrm{HCN}$ varies, the relation between HCN luminosity and dense gas mass becomes non-linear, and consequently, the \sfedense\ will have intrinsic scatter. At the spatial scales probed in this work (300\,pc), the temporal and environmental effects impact observed \sfedense, which are averaged out at larger scales.

In addition, it is important to consider the caveats of using the HCN emission as a proxy of dense gas, which has become especially important in regions with high-mean gas density. As seen in Figure\,\ref{fig:sfe_fdense}, \sfedense\, reaches a regime of $400-500\,\mathrm{L_{\odot}/M_{\odot}}$ at which \sfedense\, becomes constant, also found in previous studies \citep[e.g.,][]{shirley_2003,scoville_2004,thompson_2005,wu_2010}. In these high-density regimes, the HCN\Jone\, emission does not necessarily probe pure star-forming gas, as, for example, in other regions, where mean gas densities are comparable or lower than the critical density of HCN\Jone\, \citep{leroy_2017,beslic_2021}. Alternatively, higher $J$ transitions of HCN can better probe dense gas mass relative to the average gas density in these regions. Therefore, excited HCN emission could be used to derive the $X_\mathrm{HCN}$ factor at higher densities and constrain the dense gas mass, which is beyond the scope of this work.

\subsection{Star formation efficiency and velocity dispersion}
\label{sec:sfe_sigma}

\begin{figure}[t!]
    \centering
    \includegraphics[width=0.5\textwidth]{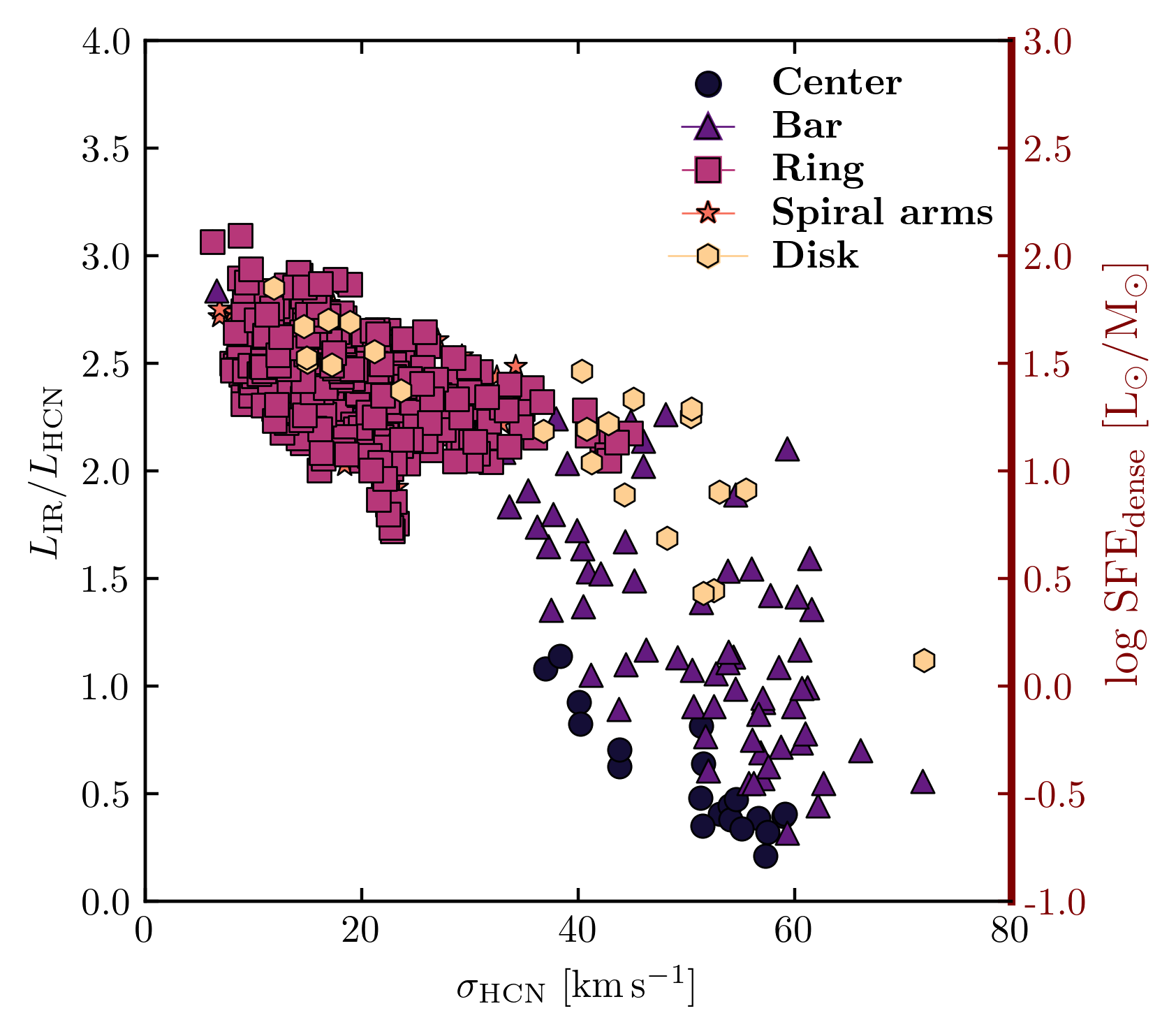}
    \caption{The ability of dense gas to form stars, traced by IR/HCN luminosity ratio, as a function of HCN velocity dispersion measured from spectral decomposition analysis. We colour-code each point by the environment.}
    \label{fig:scouse_sigma_sfedense}
\end{figure}

In this section, we investigate how the IR/HCN luminosity (and $L_\mathrm{IR}/M_\mathrm{dense}$) ratio varies with velocity dispersion and environment in NGC~253. We show the \sfedense\, as a function of the HCN velocity dispersion inferred from SCOUSE decomposition (Section\,\ref{sec:scouse}) in Figure\,\ref{fig:scouse_sigma_sfedense}. Similarly, as in Figure\,\ref{fig:scouse3}, data points populate two distinct parts of Figure\,\ref{fig:scouse_sigma_sfedense}. The first group of data points is located in the upper left part of the figure, showing HCN velocity dispersions lower than 40\,\kms, whereas the second group shows $\sigma_\mathrm{HCN}>$40\,\kms\, and lies at the bottom right part of the figure. 

The first group of data points consists of lines of sight from the ring and spiral arms, whereas the central lines of sight are in the second group. Lines of sight from the bar and the disk are found in both groups. The IR/HCN luminosity ratio decreases with increasing HCN velocity dispersion. This decrease is significantly steeper for data points with higher velocity dispersions. It is worth pointing out that the systemic rotation velocity is not necessarily resolved at our scales, which causes line broadening at the center. In addition, the observed anticorrelation between IR/HCN and $\sigma_\mathrm{HCN}$ comes from the correlation between CO and HCN intensities and their respective line widths. We note that, at the scales of our observations, the systemic rotation velocity is not fully resolved, which also broadens observed lines.

Nevertheless, broader molecular lines imply lower efficiency of such gas at star formation, and our results indicate that gas turbulence could be suppressing star formation \citep{padoan_2011, meidt_2018, meidt_2020a}. Similar results are found in M~51 and NGC~3627 \citep{querejeta_2019, beslic_2021}: central sight lines show broader emission lines, but the ability of such gas for star formation is reduced. \cite{murphy_2015} compared SFR/HCO$^+$ with HCO$^+$ line widths in the nuclear region and bar ends in NGC~3627 and concludes that the velocity dispersion of molecular is an important factor in setting star formation.


\section{Discussion}
\label{sec:discussion}
\subsection{Enhancement of the HCN/CO\Jtwo\ ratio in the inner 2\,kpc of NGC~253}
\label{sec:outflow}

In the following, we provide a brief discussion of our results presented in Sections\,\ref{sec:hcn_radial_profile} and \ref{sec:los_hcn-co}. As seen in Figure\,\ref{fig:hcn_radial_profile}, the HCN integrated intensity shows general decrease towards the higher galactocentric radii. The steepest change in the HCN intensity occurs in the inner 2\,kpc region of NGC~253. Such behaviour has been observed in previous studies probing similar or even smaller spatial scales \citep[e.g.,][]{beslic_2021,neumann_2023}. Similarly, by looking at the radial trend of HCN/CO\Jtwo\, intensity ratio (Figure\,\ref{fig:hcn_co21}), we observe change by a factor of $\sim7$ in the inner 2\,kpc, at the location of a bar in NGC~253. The reason for such steep decrease of the HCN, but also the CO\Jtwo\, intensities could be due to the bar influence. However, we speculate that other mechanisms impact this enhanced HCN/CO\Jtwo\, intensity ratio in the case of NGC~253.
The starburst in this galaxy drives the outflow seen in various gas phases: in ionized gas (H$\alpha$ and X-ray \citep{strickland_2000,strickland_2002, westmoquette_2011,lopez_2022}, neutral gas \citep{heckman_2000}, warm H$_2$ \citep{veilleux_2009}, OH both in emission and absorption \citep{turner_1985, sturm_2011}. Moreover, this outflow contains a significant amount of dust (HST observations \cite{watson_1996}), molecular gas, based on ALMA CO\Jone\ observations \citep{bolatto_2013, krieger_2020}, and dense molecular gas traced by HCN emission \citep{walter_2017, krieger_2017}.

The enhanced HCN/CO\Jtwo\ in the bar towards the centre of NGC~253 and the bimodal distribution for the bar sight lines in HCN/CO\Jtwo\ have a few possible interpretations. On the one hand, we might witness a molecular gas flow, and there is a chance that enhanced HCN/CO\Jtwo\ points in the bar originate from the molecular wind, as shown in \citet{walter_2017}. On the other hand, beam smearing effects caused by a high galaxy inclination are not negligible. Therefore, we may detect the emission originating from the centre within the bar. Moreover, \citet{paglione_2004} found high gas densities traced by CO\Jone\ emission near the inner Lindblad radius \citep[$\sim$300\,pc, ][]{iodice_2014} and along the bar's minor axis. We investigate these sight lines to further explore and explain the enhancement in HCN/CO\Jtwo. 

\begin{figure*}[t!]
        \includegraphics[width=\textwidth]{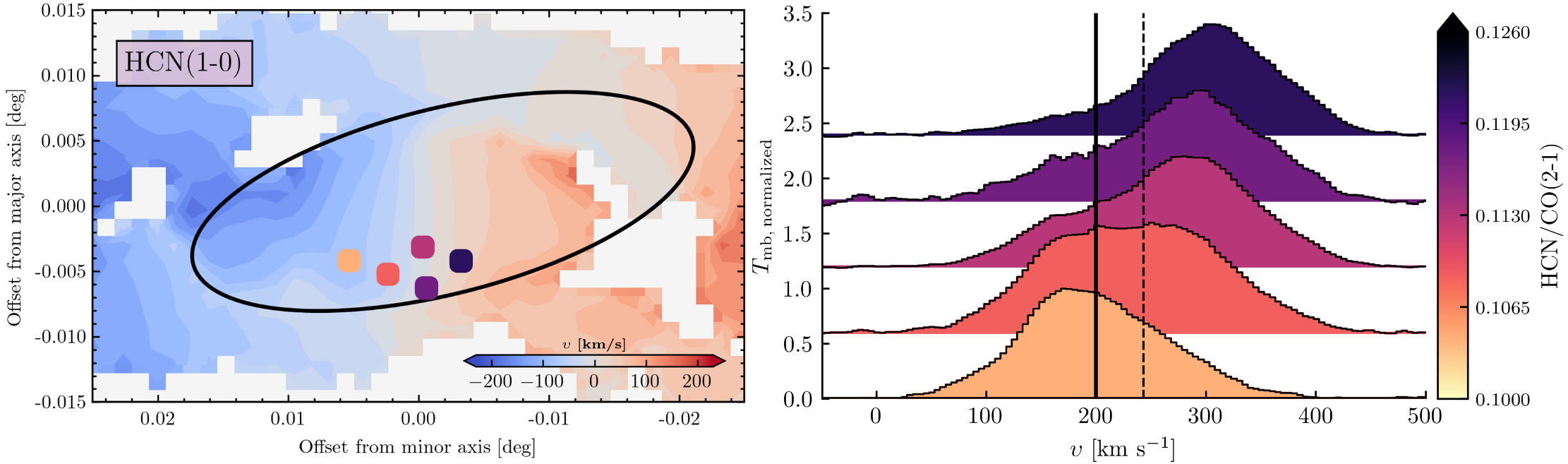}
	\caption{\textbf{Left:} Bar sight lines in NGC~253. We label major and minor axes. Coloured points are sight lines showing enhanced HCN/CO\Jtwo\ emission in the direction of the SW streamer \citep{walter_2017} and that contribute to the bimodality seen in Figure\,\ref{fig:hcn_co21}. Colors correspond to their HCN/CO\Jtwo\ intensity ratio shown in the colorbar. \textbf{Right:} HCN spectra towards the data points located along the minor axis in NGC~253 shown on the left. Each spectrum is coloured by the HCN/CO\Jtwo\ ratio shown on the colourbar. All these data points are located along the SW streamer \citep{walter_2017}. The black dashed line shows NGC~253 systemic velocity of $243$\,km\,s$^{-1}$ (see Table\,\ref{tab:NGC0253_prop}). The black solid line represents the $\upsilon_\mathrm{lsr}=200$\,km\,s$^{-1}$ velocity at which \citet{walter_2017} observed the outflowing component in CO\Jone\ emission. }
    \label{fig:hcn_spectra}
\end{figure*}

The bar's data points that form the upper locus of HCN/CO\Jtwo\ (right panel of Figure\,\ref{fig:hcn_co21}) have values above 0.1. After inspecting the data points that cause the upper locus in the HCN/CO\Jtwo, we find that these points lie along the minor axis in NGC~253, i.e., where we expect to see the outflow. It is worth noting that the spatial scales probed by our data are not sufficient to resolve the outflow. Nevertheless, it is still possible to see the signs of the outflow within the spectra. \citet{walter_2017} observed the wind in NGC~253 in CO\Jone\ emission and found a separate velocity component at intermediate ($-4$" to $-10$") offsets from the major axis. Therefore we plot the HCN spectrum including the decomposed emission derived from SCOUSE (Section\,\ref{sec:scouse}) towards the sight lines with enhanced HCN/CO\Jtwo\ found along the minor axis, towards the direction of the brightest, south-western (SW) streamer of the outflow (negative offsets from the minor axis). We show these in Figure\,\ref{fig:hcn_spectra}. Each spectrum is coloured by the HCN/CO\Jtwo\ line intensity ratio. 

Across all these points, we observe that the HCN emission consists of two components, the dominant one (dashed line) centered at velocities higher than 243\,km\,s$^{-1}$, and a second one (solid line) centered at velocities lower than $-43\,$km\,s$^{-1}$, consisted with the value for the outflow, $\upsilon=-43$\,km\,s$^{-1}$ \citep{walter_2017}. We show this velocity as a solid black line in Figure\,\ref{fig:hcn_spectra}, and systemic NGC~253 velocity as a black dashed line.

\citet{walter_2017} calculated HCN/CO\Jone\ line intensity ratio in the outflow and the disk of NGC~253. This study found that the HCN/CO\Jone\ is $\sim$0.1 in the outflow, and three times lower value in the disk ($1/30$). We estimate the HCN/CO\Jone\ ratio from our measurements. To do so, we use a mean $R_{21}$ ratio of 0.8 measured in NGC~253 \citep{zschaechner_2018}. After applying the $R_{21}$ ratio, sight lines with enhanced HCN/CO\Jtwo\ line intensity ratio (from 0.1 to 0.12) correspond to values within the 0.080 to 0.096 range in HCN/CO\Jone. Therefore, the molecular outflow could explain the enhanced HCN/CO\Jtwo\ intensity ratio found along the minor axis. 

\subsection{Molecular gas kinematics}
\label{sec:kinematics}

The inner 500\,pc region of NGC~253, also called the central molecular zone (CMZ) is characterized by a strong dust continuum emission \citep{leroy_2018}. The giant molecular clouds found in the CMZ of NGC~253 \citep{leroy_2015} are more massive and have higher velocity dispersion than the GMCs in the Milky Way \citep{krieger_2020}. These clouds contain young massive clusters, bright in continuum and line emission \citep{leroy_2015, leroy_2018}, with detected large-scale outflow in dense molecular \citep{levy_2021} and ionized gas \citep{mills_2021}. 

\cite{Meier_2015} provided a schematic representation of the central region in NGC~253, demonstrating its interesting and complex structure. These authors found that the nuclear disk consists of an inner and outer region ($\sim170$\,pc and $\sim400$\,pc). The inner nuclear disk contains high-density molecular gas and intense star formation. In contrast, the outer nuclear disk is considered the region where gas flows inward along the large-scale bar \citep{sorai_2000, paglione_2004, Meier_2015}.

\begin{figure*}
    \centering
    \includegraphics[width=\textwidth]{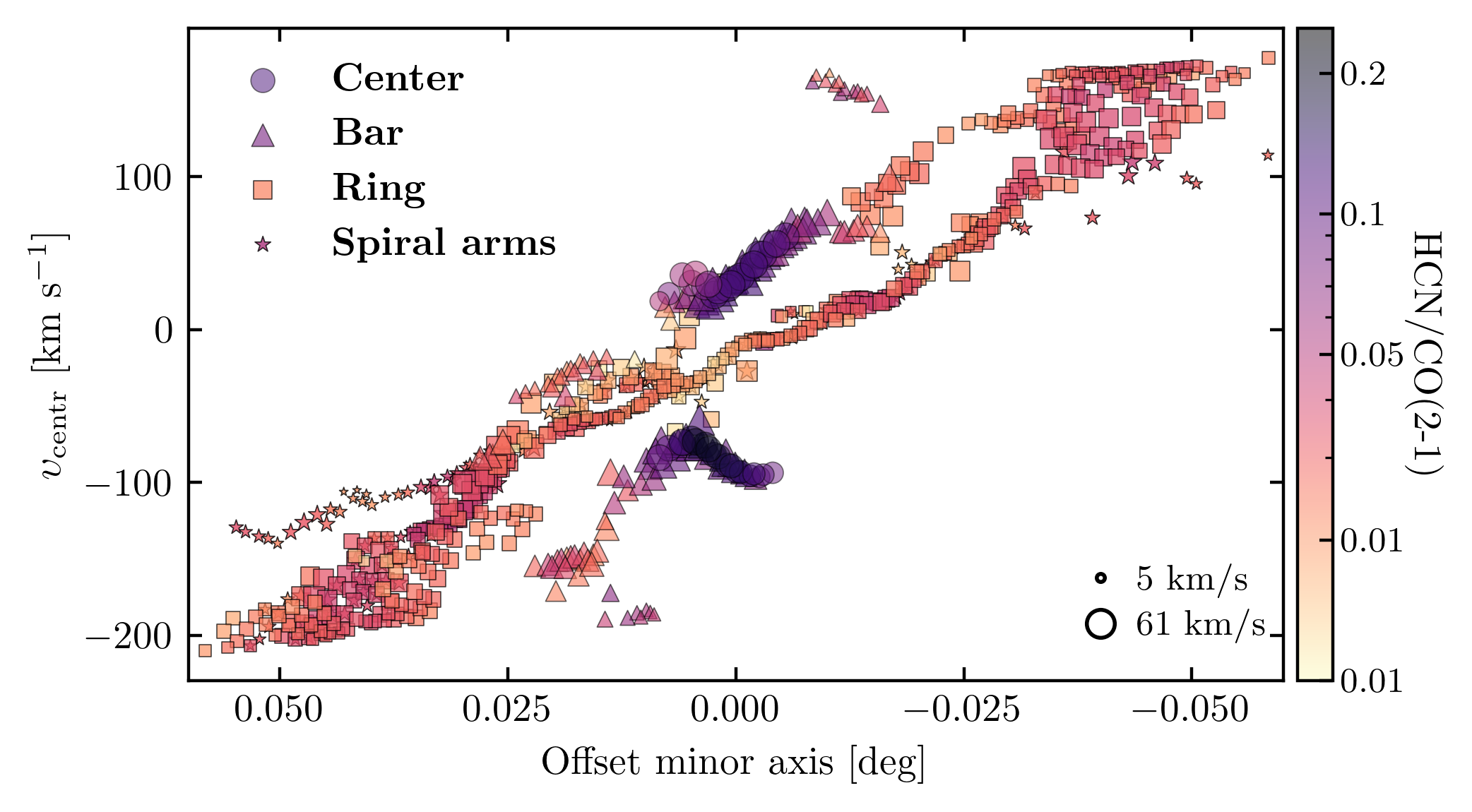}
    \caption{Position-velocity diagram (p$\upsilon$) of HCN emission in NGC~253. Y-axis are the centroid velocities from the spectral decomposition, and on the x-axis are angular distances from the minor axis of NGC~253. We colour-code points by their HCN/CO\Jtwo\ ratio derived from SCOUSE, and the size of each point corresponds to the HCN velocity dispersion.}
    \label{fig:scouse_pvd}
\end{figure*}

We show the position-velocity diagram in Figure\,\ref{fig:scouse_pvd} of the centroid velocities of HCN emission from the SCOUSE as a function of the offset from the galaxy's minor axis. Each point is colour-coded by the HCN/CO\Jtwo\ integrated intensity and the environment. The HCN/CO\Jtwo\ is derived from the integrated intensities calculated from SCOUSE (see Section\,\ref{fig:scouse_hcn_co21_ratio}). Bar sight lines tend to form a "parallelogram"-shaped feature in the pv plane, previously seen in CO\Jone\ \citep{sorai_2000} and CS\Jtwo\ emission \citep{peng_1996}, consistent with gas flowing in a bar potential \citep{binney_1991, athanassoula_1992, peng_1996, sormani_2015a}. We note that there are more data points on the leading side of the bar (bottom left part of Figure\,\ref{fig:scouse_pvd}), which is indicative of molecular gas concentrations caused by a non-axisymmetric bar potential \citep{kuno_2000}. The velocity gradient is the steepest in the bar and centre and more shallow in the ring and spiral arms. As the molecular gas travels from the outer disc, it starts losing angular momentum at galactocentric radii lower than the co-rotation radius $\sim3{-}4$\,kpc \citep[e.g.][]{iodice_2014}, and is accreted onto the ring. The rotation curve within the ring is almost flat, causing the crowding of molecular gas, which is possibly supported by the observed narrow single-peaked HCN spectral features.

\subsection{Star formation efficiency}
\label{sec:sfe}

\begin{figure*}[t!]
\centering
    \includegraphics[width = \textwidth]{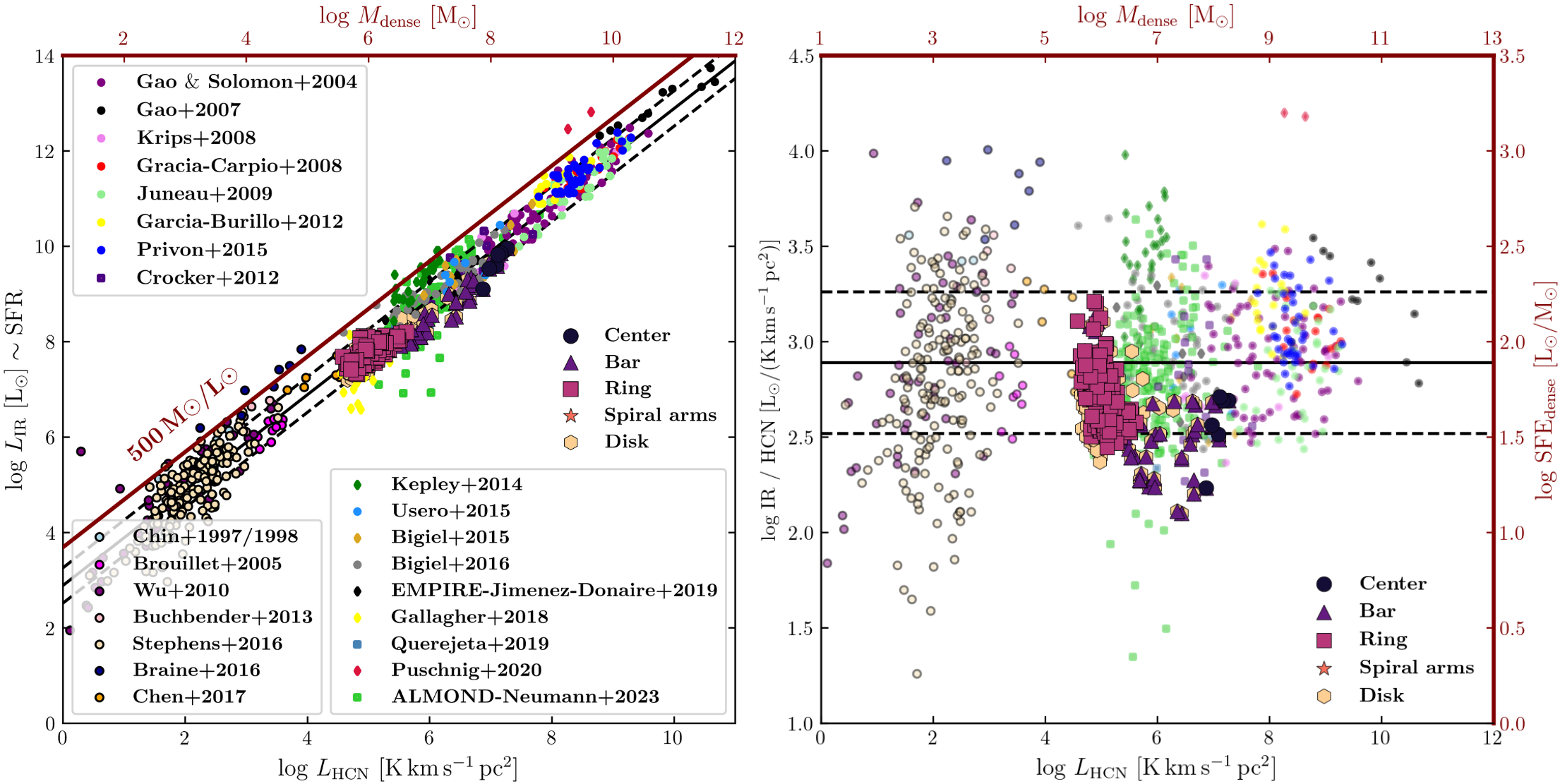}
    \includegraphics[width = \textwidth]{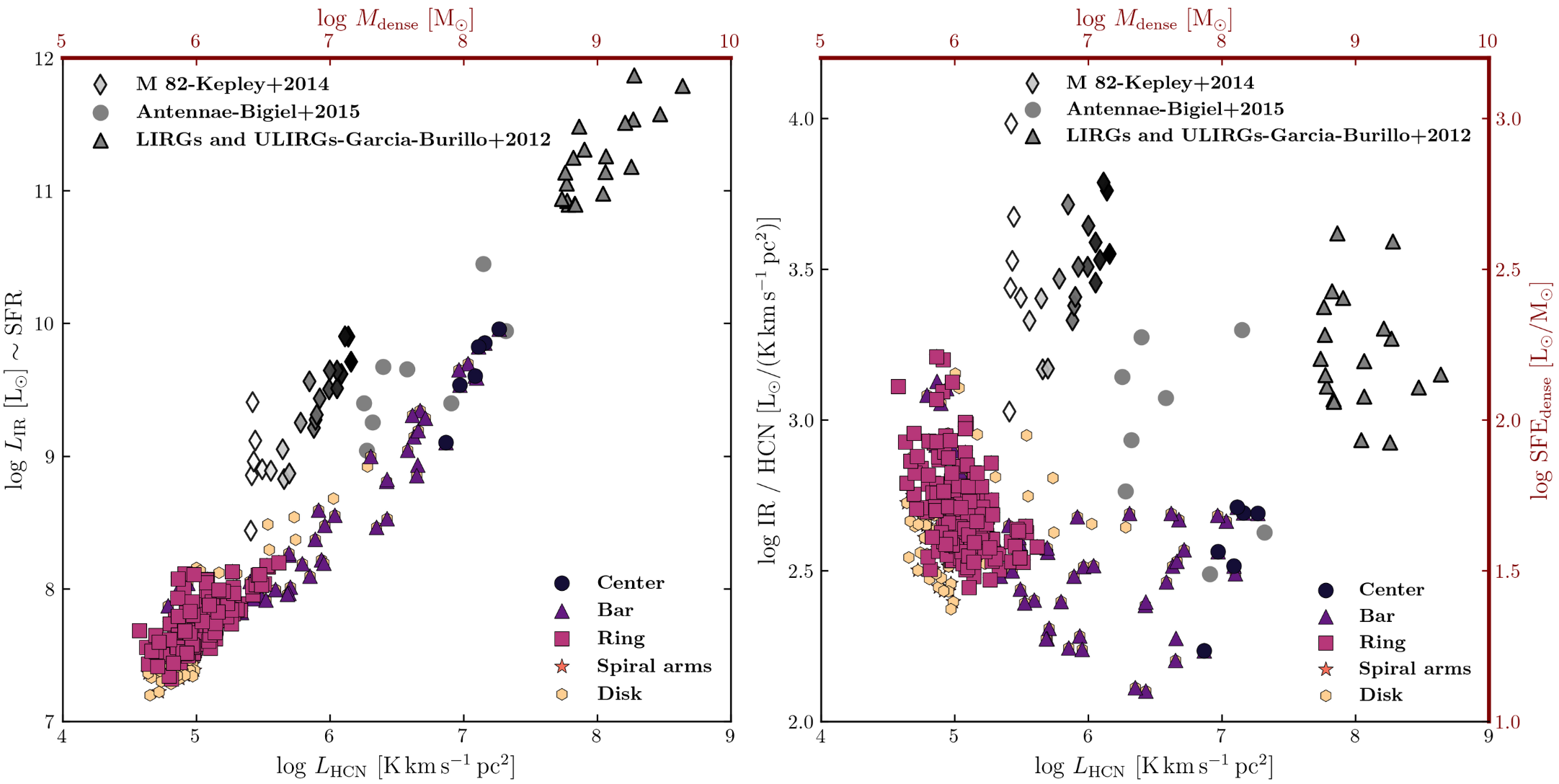}
    \caption{\textbf{Top Left:} A literature overview of the infrared luminosity (a proxy for star formation rate) as a function of HCN luminosity (a proxy for dense molecular gas), i.e.\ Gao \& Solomon relation \citep{gao_2004a}. We show measurements of whole galaxies and their centres at a few kpcs (see the legend in the upper and the lower-left of the panel), Milky Way clouds (a few pc scales - see legend on the bottom right), including measurements across NGC~253 (this work - see the legend on the right central part) \citep{jimenez19}. The dark red line shows a $500\,$M$_{\odot}$/L$_{\odot}$ which corresponds to a upper limit reported in \cite{scoville_2004,thompson_2005}. \textbf{Top Right:} Same as on the left panel, but we are now showing SFE$_\mathrm{dense}$, traced by the IR/HCN luminosity ratio on the y axis. In both panels, the black solid line shows the mean IR/HCN luminosity ratio measured by the EMPIRE survey \citep{jimenez19}. Black dashed lines show the 1$\sigma$ RMS scatter of $\pm$0.37\,dex in units of $\mathrm{L_{\odot}/(K\,km\,s^{-1}\,pc^{2})}$. \textbf{Bottom Row:} A zoom-in field from Figure\,\ref{fig:gs}, where we show NGC~253 sight lines, points from the nearby starburst galaxy M~82 \citep{kepley_2014}, the merging system of the Antennae galaxies \citep{bigiel_2015}, and a sample of LIRGs and ULIRGs \citep{garcia_burillo_2012}. Axes are the same as of the panels at the top row. Data points from \citet{kepley_2014} have different shades of grey that corresponds to the distance to the centre of M~82, where the darkest points have the smallest galactocentric distances.}
    \label{fig:gs}
\end{figure*}

In Figure\,\ref{fig:gs}, we summarize previous HCN surveys and studies, including recent ones \citep[e.g. results from the ALMOND survey][]{neumann_2023}. We show the relation of HCN luminosity and dense gas mass relation to star formation on the top left panel of Figure\,\ref{fig:gs}, and to the TIR/HCN luminosity ratio and $L_\mathrm{IR}/M_\mathrm{dense}$ \citep[top right panel, see also][]{jimenez19}.

We observe an approximately linear correlation between IR and HCN luminosities. As seen from the top panels of Figure\,\ref{fig:gs}, the literature overview shows this correlation spanning more than ten orders of magnitude, covering a wide range of physical scales, from dense clumps and cores \citep[a few pc - e.g.,][]{wu_2010, stephens_2016} within the Milky Way, to GMCs in local and nearby galaxies \citep{chin_1997,chin_1998,braine_2017,brouillet_2005,buchbender_2013,chen_2017,querejeta_2019}, measurements of resolved galaxy disks (a few kpcs \citealp{kepley_2014,usero_2015,bigiel_2016,chen_2015,gallagher_2018a,tan_2018,jimenez19,jiang_2020}) to whole galaxies and their centers \citep{gao_2004a,gao_2007,krips_2008,gracia_carpio_2008,juneau_2009,garcia_burillo_2012,crocker_2012,privon_2015, puschnig_2020}. By observing this trend, a direct conclusion is that more dense molecular gas implies a higher star formation rate.


Therefore, we investigate how our results fit the relationship between star formation and dense molecular gas within the literature. We also include measurements from this work on NGC~253, colored by different regions in this galaxy in Figure\,\ref{fig:gs}.

Our measurements span $\sim$ 3 orders of magnitude in IR luminosity and dense gas mass. Lines of sight from NGC~253 fall into the same region of the SFR-HCN plane as measurements of whole galaxies and parts of galaxies. The center of NGC~253 contains the largest amount of dense gas and shows the highest star formation activity, unlike the rest of the NGC~253's environments. By contrast, in the top right panel, we note that regions with the brightest HCN emission in NGC~253 have the lowest \sfedense. This is in agreement with previous studies. For instance, the CMZ, the densest and most active region in our Galaxy, shows an order of magnitude lower SFR than those predicted from measurements of dense molecular gas \citep{longmore_2013a, henshaw_2022}. In extragalactic work, \cite{jimenez19,neumann_2023} observed reduced \sfedense\, towards centers of nearby galaxies, despite containing most of $M_\mathrm{dense}$ and the highest \fdense. \cite{querejeta_2019} have shown that gas in spiral arms has large \sfedense\, in M~51. Similarly, bar ends in NGC~3627 show higher \sfedense\, than the center of this galaxy \citep{beslic_2021}.

To first order, this relation seen in top panels of Figure\,\ref{fig:gs} suggests a universal density threshold above which gas starts to form stars \citep{lada_2012}. However, the systematic scatter in the right panel in Figure\,\ref{fig:gs} implies that not all dense gas is equally efficient at star formation, i.e., \sfedense\, is environmentally dependent. On the one hand, this could imply a physical change in $L_\mathrm{IR}/M_\mathrm{dense}$, possibly driven by dynamical effects such as turbulence and stellar feedback \citep[e.g.,][]{padoan_2011,federrath_2012} that become important at GMC scales \citep{gallagher_2018a, querejeta_2019, sanchez-garcia_2022, neumann_2023}. In addition, as previously discussed in Section\,\ref{sec:sfe_fdense}, a systematic variation in $X_\mathrm{HCN}$ can cause the observed scatter.

On the other hand, even in the case of a constant conversion factor between HCN luminosity and dense gas mass, it is possible that HCN might not be a reliable linear tracer of $M_\mathrm{dense}$, resulting in not all HCN-tracing gas taking part in star formation.  

In the bottom panels of Figure\,\ref{fig:gs}, we compare our results with other HCN observations across other starburst systems. We show measurements from \citealp{kepley_2014} across the nearby starburst galaxy M~82 at slightly smaller physical scales (200\,pc) than our work the sample of ULIRGs from \citealp{garcia_burillo_2012}, and sight lines from the galaxy merger, Antennae at 700\,pc \citep{bigiel_2015}. LIRGs and ULIRGs show elevated IR and HCN luminosity, two orders of magnitude higher than starburst systems. M~82 sight lines appear to have higher IR luminosity than NGC~253. The Centre of M~82 exhibits the brightest HCN emission \citep[grey diamonds in the bottom row of Figure\,\ref{fig:gs}][]{kepley_2014}, although it is fainter than the HCN emission towards the center of NGC~253. M~82 and NGC~253 are both typical starburst galaxies. However, these galaxies go through a different evolutionary phase \citep{rieke_1988} and have different magnitudes of starburst. NGC~253 is in the earlier starburst stage than M~82 \citep[$\sim10$\,M$_{\odot}$~yr$^{-1}$][]{telesco_1991}. In addition, M~82 is a gas-rich starburst triggered by a major interaction with M~81. Therefore, we expect a higher fraction of HCN-tracing gas across NGC~253. 

Looking at the top right panel in Figure\,\ref{fig:gs}, we note scatter in the IR/HCN ratio of $\sim$3\,dex for all sources. Overall, among each source presented in this figure, we observe that lines of sight with the highest HCN luminosities have the lowest star formation efficiencies, except M~82, where this trend is unclear. LIRGS and ULIRGS \citep{garcia_burillo_2012}, together with measurements of M~82 \citep{kepley_2014}, have the highest star formation efficiencies. 

The results discussed in this section are consistent with the known role of the HCN-tracing gas in star formation, i.e., that higher HCN luminosity implies higher star formation rate, but that the scatter in IR/HCN as a function of HCN luminosity is not negligible. We observe environmental dependence on the observed IR/HCN with dense gas mass, as confirmed in several previous studies \citep[e.g.][]{jimenez19}. However, The still observed scatter remains to be explained by future more sensitive, high-resolution observations across nearby galaxies that can probe the physical properties of individual molecular clouds.


\section{Summary}
\label{sec:summary}
We present new ALMA ACA+TP observations of HCN emission at 300\,pc scales across the closest southern starburst galaxy NGC~253. These observations cover a large portion of NGC~253 disk that contains 95$\%$ of detected CO\Jtwo\ emission obtained by ALMA ACA, and 85$\%$ of the star formation activity measured from ancillary infrared data at 70, 160, and 250 $\mu$m obtained by Herschel Space Telescope \citep{pilbratt_2010}. Our work investigates the HCN line intensity distribution, its relation to the CO\Jtwo\ emission, gas kinematics traced by this molecular line, the ability of gas to form stars, and the environmental dependence of these properties. Here we summarize our results:

\begin{itemize}
    \item We use two independent methods to derive the integrated intensity of HCN emission. The first approach is based on calculating the zeroth moment map and extracting the information of HCN emission from it. In contrast, the second one uses information derived from the spectral decomposition of the observed emission along each line of sight. By performing spectral decomposition, we gain insight into the velocity components that contribute to the observed line emission at each point. \\
    \item Our results derived from both methods are in good agreement. We find environmental differences in the observed HCN emission across NGC~253, particularly in regions closer to the center and regions in the disc. The HCN emission is strongly enhanced towards the center of NGC~253, and its intensity decreases by two orders of magnitude along the bar. HCN intensity weakly varies along the ring, spiral arm, and disc. \\
    \item In addition, the HCN spectrum shows complexity and environmental dependence. In the inner 2\,kpc region, we observe multiple velocity structures in the HCN and CO\Jtwo spectra. Using spectral decomposition, we find up to three separate velocity structures in the central and bar sight lines, indicating that our observations probe multiple gas flows. Moreover, HCN and CO\Jtwo\ spectral lines within the ring structure and spiral arms are mainly single-peaked and significantly narrower than the spectra from the inner regions. Our results support the idea that the molecular gas sits in the ring, after which it is inflowing towards the nuclear region along the bar.
    \item We investigate HCN/CO\Jtwo\ line intensity ratio and its environmental distribution in NGC~253. We found that the HCN/CO\Jtwo\ distribution within the bar shows bimodal behavior, which is a consequence of beam-smearing effects and a possible indication of the molecular outflow found in \citet{walter_2017}. A nearly constant HCN/CO\Jtwo\ intensity ratio within the ring, in line with the flattened rotational curve of CO at this region, suggests that molecular gas traced by HCN emission gets piled up in that region. \\
    \item The majority of the decomposed HCN emission has an associated CO\Jtwo\ component. All lines of sight in the ring, spiral arms, and disc have HCN emission associated with the CO\Jtwo. In contrast, in inner regions, the percentage of associated components is somewhat lower, which could be due to opacity broadening, lower signal-to-noise ratio in HCN than CO\Jtwo\ emission, but also that some of the CO\Jtwo\ is diffuse and does not contain HCN. Together with spectral complexity, we find the environment to be the driver of the measured HCN velocity dispersion and broader HCN lines in regions with different dynamical characteristics. \\
    \item Wider CO\Jtwo\ profiles contain wider HCN emission lines, and CO\Jtwo\ emission lines are broader than HCN lines, suggesting that HCN-tracing gas arises from smaller spatial structures. This result aligns with turbulence theory, which predicts that high-density gas is mostly seen at the stagnation regions of larger-scale convergent flows. \\
    \item We investigate the ability of the gas to form stars as a function of the relative amount of HCN-traced molecular gas to bulk molecular gas. Lines of sight with the highest HCN/CO\Jtwo\ ratio appear the least efficient at star formation (the lowest IR/HCN). This result suggests that the processes that regulate star formation vary with the environment.
\end{itemize}

Our work illustrates the importance of mapping the spatially resolved emission of the molecular gas tracing star-forming content on the example of NGC~253. In this work, we also analyze the spectrum of these lines along individual lines of sight. We highlight the necessity of extending our understanding of dense gas properties across starburst systems such as NGC~253 since they provide additional constraints on the dense gas properties in the vicinity of extreme physical conditions, particularly in its center.

\begin{acknowledgements}
We thank the anonymous referee for their insightful comments that improved this manuscript. The author wants to thank Maryvonne Gerin and Jerome Pety for their insightful discussion that contributed to the quality of this work.

This paper makes use of the following ALMA data: ADS/JAO.ALMA\#2019.2.00236.S and ADS/JAO.ALMA\#2018.1.01321.S. ALMA is a partnership of ESO (representing its member states), NSF (USA) and NINS (Japan), together with NRC (Canada), MOST and ASIAA (Taiwan), and KASI (Republic of Korea), in cooperation with the Republic of Chile. The Joint ALMA Observatory is operated by ESO, AUI/NRAO and NAOJ. In addition, publications from NA authors must include the standard NRAO acknowledgement: The National Radio Astronomy Observatory is a facility of the National Science Foundation operated under cooperative agreement by Associated Universities, Inc.

IB, ATB, and FB would like to acknowledge the funding provided from the European Union's Horizon 2020 research and innovation programme (grant agreement No 726384/Empire).
HAP acknowledges support by the National Science and Technology Council of Taiwan under grant 110-2112-M-032-020-MY3.
KG is supported by the Australian Research Council through the Discovery Early Career Researcher Award (DECRA) Fellowship DE220100766 funded by the Australian Government. KG is supported by the Australian Research Council Centre of Excellence for All Sky Astrophysics in 3 Dimensions (ASTRO~3D), through project number CE170100013.
MC gratefully acknowledges funding from the DFG through an Emmy Noether Research Group (grant number CH2137/1-1).
COOL Research DAO is a Decentralized Autonomous Organization supporting research in astrophysics aimed at uncovering our cosmic origins.
JMDK gratefully acknowledges funding from the European Research Council (ERC) under the European Union's Horizon 2020 research and innovation programme via the ERC Starting Grant MUSTANG (grant agreement number 714907). 
RSK acknowledges support from DFG via  the collaborative research center ``The Milky Way System'' (SFB 881; project ID 138713538; sub-projects B1, B2 and B8), from the Heidelberg cluster of excellence EXC 2181``STRUCTURES'' (project ID 390900948), funded by the German excellence strategy, from ERC via the synergy grant ``ECOGAL'' (grant 855130), and from the German Ministry for Economic Affairs and Climate Action for funding in  project ``MAINN'' (funding ID 50OO2206).

\end{acknowledgements}


\bibliographystyle{aa}
\bibliography{references.bib} 

\begin{appendix}
\section{Molecular surface density}
\label{sec:app-smol}

\begin{figure*}[t!]
    \centering
    \includegraphics[width=\textwidth]{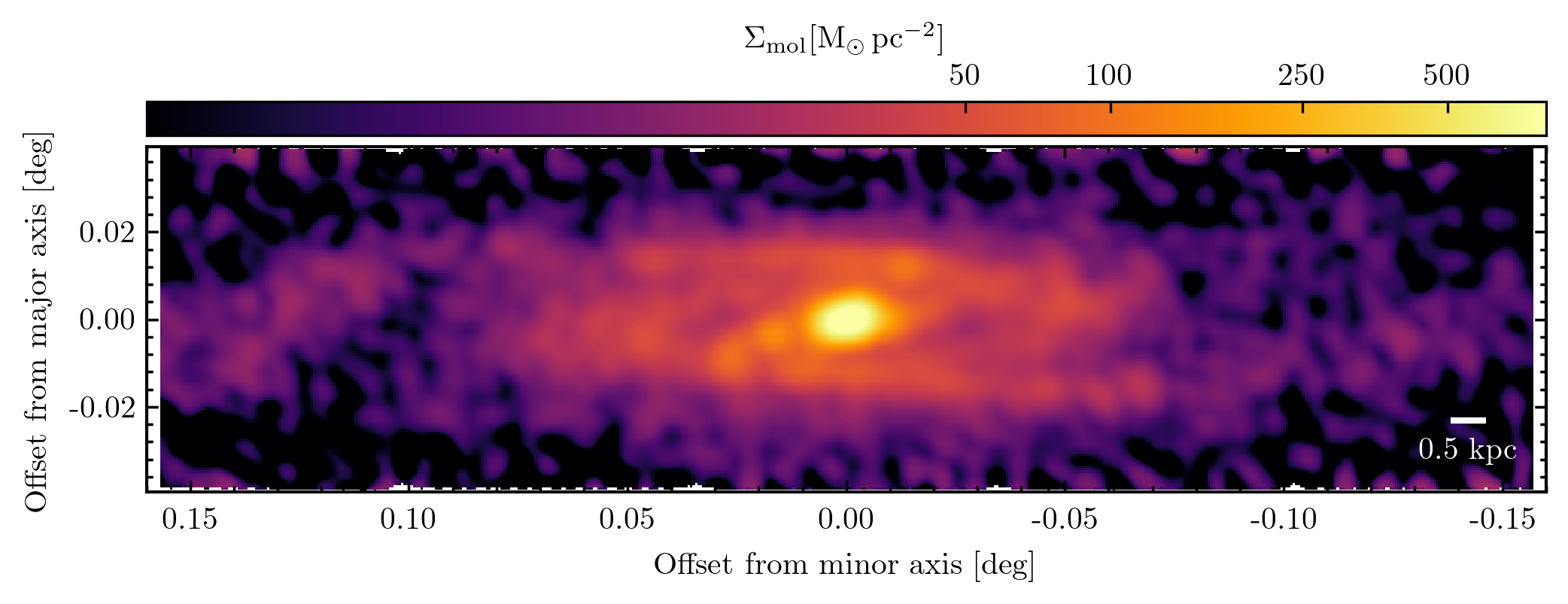}
    \caption{Molecular surface density map of NGC~253 derived from CO\Jtwo\, emission.}
    \label{fig:smol_map}
\end{figure*}

We show \sigmamol\, in Figure\,\ref{fig:smol_map}. The production of this map is described in Section\,\ref{sec:smol}.  In this work, we used a specific region of CO\Jtwo\ emission, highlighted by the dashed white rectangle in Figure\,\ref{fig:ngc253}.

\section{Star formation surface density}
\label{sec:app-sfr}

\begin{figure*}[h!]
    \centering
    \includegraphics[width=\textwidth]{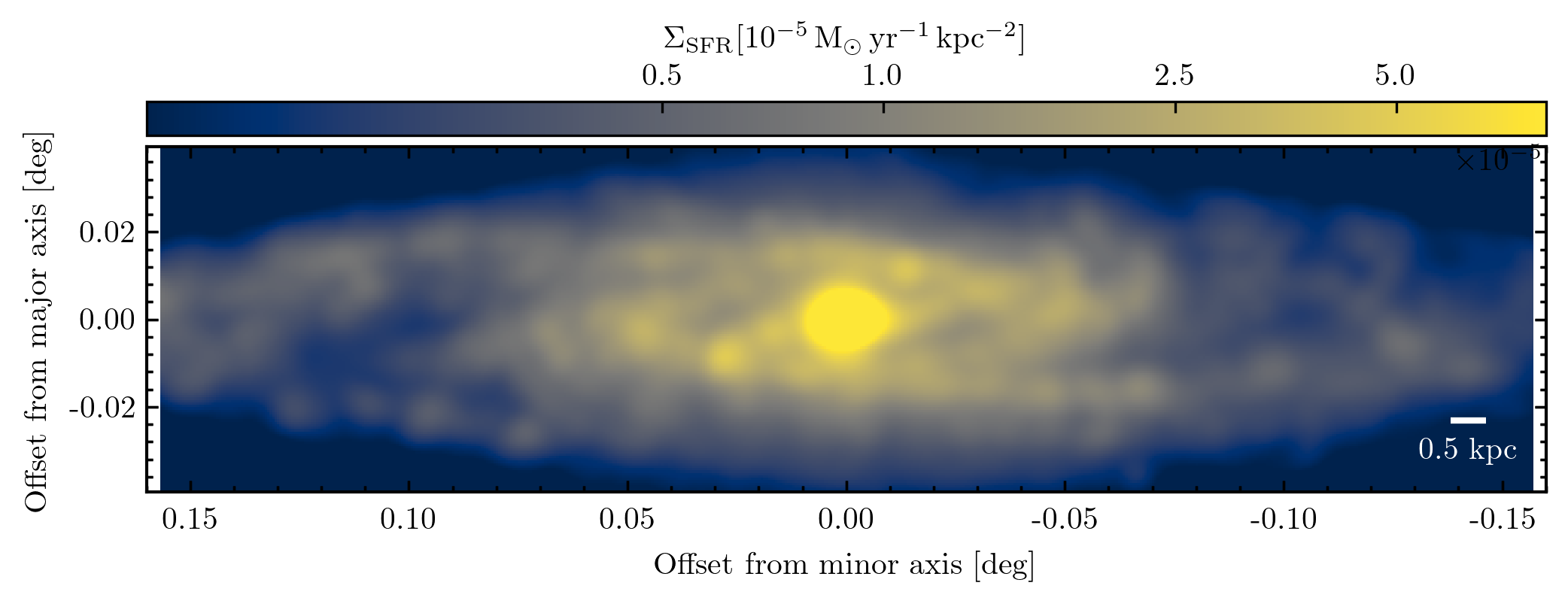}
    \caption{Star formation surface density map of NGC~253 using Herschel data.}
    \label{fig:sfr_map}
\end{figure*}

In Table\,\ref{tab:coeff_tir} we show coefficients used to compute the star formation surface density, which map we show in Figure\,\ref{fig:sfr_map}.

\begin{table}[h!]
\centering
\caption{Coefficient for calculation total infrared surface density (Equation\,\ref{eq:tir}, Section\,\ref{sec:ssfr}) taken from \cite{galametz_2013}.}
\begingroup
\setlength{\tabcolsep}{8pt} 
\renewcommand{\arraystretch}{1.2} 
\begin{tabular}{cc}
\hline \hline
$j=\lambda$ {[}nm{]} & $c_j$             \\ \hline
70                   & 1.018 $\pm$ 0.021 \\
160                  & 1.068 $\pm$ 0.035 \\
250                  & 0.402 $\pm$ 0.097 \\ \hline \hline
\end{tabular}
\endgroup
\label{tab:coeff_tir}
\end{table}

\section{Spectral decomposition using SCOUSE}
\label{sec:app-scouse}

In our work, we use the Semi-automated multi-COmponent Universal Spectral-line fitting Engine (SCOUSE) \citep{henshaw_2016a, henshaw_2019}, which is the spectral decomposition algorithm that defines where the line is located within the spectrum and describes such lines using Gaussian fitting. SCOUSE consists of four steps. In the first step, we define a spatial area over which we want to fit the spectra by creating a grid of macro pixels called Spectral Averaging Areas (SAAs). The size of a SAA is a free parameter. In our case, each SAA contains 4 pixels. The spectrum of each SAA represents the averaged spectrum of all the spectra within pixels contained in the SAA. In the second phase, SCOUSE fits each SAA's spectrum and suggests a model solution assuming the Gaussian line profile. The model solution within each SAA consists of the detected number of Gaussian components and the fitting parameters: the amplitude, the centroid velocity, and the line width. In case when SCOUSE cannot find a proper solution for a fit, the user can manually change the requested signal-to-noise ratio in order for fitting parameters to converge or fit the respective spectrum manually by selecting the location of the spectral line within the spectrum, its peak, and the brightness temperature at which the spectrum reaches the 50\,$\%$ of its peak. After completing the second phase, SCOUSE has a model solution of each SAA's spectrum and uses these to fit pixels within each SAA. In the third phase, the user checks the SCOUSE's model solutions for each pixel. Like the second phase, the user can change the number of Gaussian components and their fitting parameters. The location of each pixel in the map and SCOUSE solutions for their spectrum are saved in the textual file in the final, fourth phase.
\end{appendix}
\end{document}